\documentclass{iopart}
\usepackage{graphicx}
\usepackage{amssymb}
\usepackage{color}
\usepackage{listings}
\usepackage{cite}
\usepackage{hyperref}

\usepackage{tikz}
\usetikzlibrary{matrix}

\newcommand{\dotprod}[2]{\langle #1, #2 \rangle}
\newcommand{\eqref}[1]{(\ref{#1})}

\begin{document}

\newlength\figureheight 
\newlength\figurewidth

\topical{The instanton method and its numerical implementation in fluid mechanics}

\author{Tobias Grafke$^1$, Rainer Grauer$^2$ and Tobias Sch\"afer$^3$}
\address{$^1$ Department of Physics of Complex Systems, Weizmann Institute, Rehovot, Israel}
\address{$^2$ Theoretische Physik I, Ruhr-Universit\"at Bochum, Germany}
\address{$^3$ Department of Mathematics, College of Staten Island, Staten Island, USA \& \\ 
               Physics Program at the CUNY Graduate Center, New York, USA}
\eads{\mailto{tobias.grafke@weizmann.ac.il}, \mailto{grauer@tp1.rub.de}, \mailto{tobias@math.csi.cuny.edu}}

\date{\today}

\begin{abstract}
  A precise characterization of structures occurring in turbulent
  fluid flows at high Reynolds numbers is one of the last open
  problems of classical physics. In this review we discuss recent
  developments related to the application of instanton methods to
  turbulence. Instantons are saddle point configurations of the
  underlying path integrals. They are equivalent to minimizers of the
  related Freidlin-Wentzell action and known to be able to
  characterize rare events in such systems. While there is an
  impressive body of work concerning their analytical description,
  this review focuses on the question on how to compute these
  minimizers numerically. In a short introduction we present the
  relevant mathematical and physical background before we discuss the
  stochastic Burgers equation in detail. We present algorithms to
  compute instantons numerically by an efficient solution of the
  corresponding Euler-Lagrange equations. A second focus is the
  discussion of a recently developed numerical filtering technique
  that allows to extract instantons from direct numerical
  simulations. In the following we present modifications of the
  algorithms to make them efficient when applied to two- or
  three-dimensional fluid dynamical problems.  We illustrate these
  ideas using the two-dimensional Burgers equation and the
  three-dimensional Navier-Stokes equations.
\end{abstract}

\maketitle

\section{Introduction}

Using saddle point techniques in statistical and quantum physics has a
long history since the early work related to disordered systems in
solid state physics
\cite{lifshitz:1964,halperin-lax:1966,zittartz-langer:1966,langer:1967,langer:1969}.
Especially, the pioneering works of Zittartz and Langer
\cite{zittartz-langer:1966,langer:1967,langer:1969} contain all the
ingredients that are relevant for this review.
The term ``instanton'' was introduced in Yang-Mills theory as the
classical solution of equations of motion in the Euclidean space with
nontrivial topology in \cite{belavin-polyakov-schwartz-etal:1975}. The
quantum effects of instantons (determinant of quantum fluctuations)
were computed for the first time by 't~Hooft in \cite{thooft:1976}.
The main features of the instanton calculus as a non-perturbative
method for the calculation of path integrals were highlighted on the
example of 0+1 quantum mechanical systems in excellent reviews
\cite{coleman:1979,vainshtein-zakharov-etal:1982}. The instanton calculus 
consists basically of four steps:

\begin{enumerate}
  \item Calculation of the instanton as a classical trajectory (minima
    of the corresponding action $S$): the instanton provides the
    exponential decay term $\exp(-S)$ in the transition
    amplitude. \label{itm:instanton-calculation}
  \item Calculation of \emph{zero modes} that leave the action
    invariant: finding the zero modes is closely related to finding
    the symmetries of the underlying system. Once the zero modes are
    determined and if their number $p$ is finite, as is usually the
    case, their contribution results from a finite dimensional
    integral and often takes the form
    $(\sqrt{S})^p$. \label{itm:zero-modes}
  \item Calculation of the path integral of \emph{fluctuations} around the
    instanton which change the action: this is normally done in the
    Gaussian approximation.
  \item Summation over the \emph{instanton gas}. \label{itm:gas}
\end{enumerate}

Each of these steps may involve major obstacles but the instanton
calculus offers an algorithmic approach for the calculation of
transition probabilities in quantum mechanical and statistical
systems, including the exponential decay and leading power law
scaling. In this report we will focus on step
(\ref{itm:instanton-calculation}): calculation of the instanton. We
will shortly discuss the obstacles for the next steps and outline
first ideas.  However, already the knowledge of the instanton provides
a deep insight into the underlying system which e.g. in the context of
rare fluctuations is the adequate answer. One should also note that
the expressions related to steps
(\ref{itm:zero-modes})--(\ref{itm:gas}) are (complicated) functionals
of the instantons. Thus, step (\ref{itm:instanton-calculation}) is
naturally the most important step in this analysis.

\begin{figure}
  \begin{center}
    \includegraphics[width=0.48\textwidth]{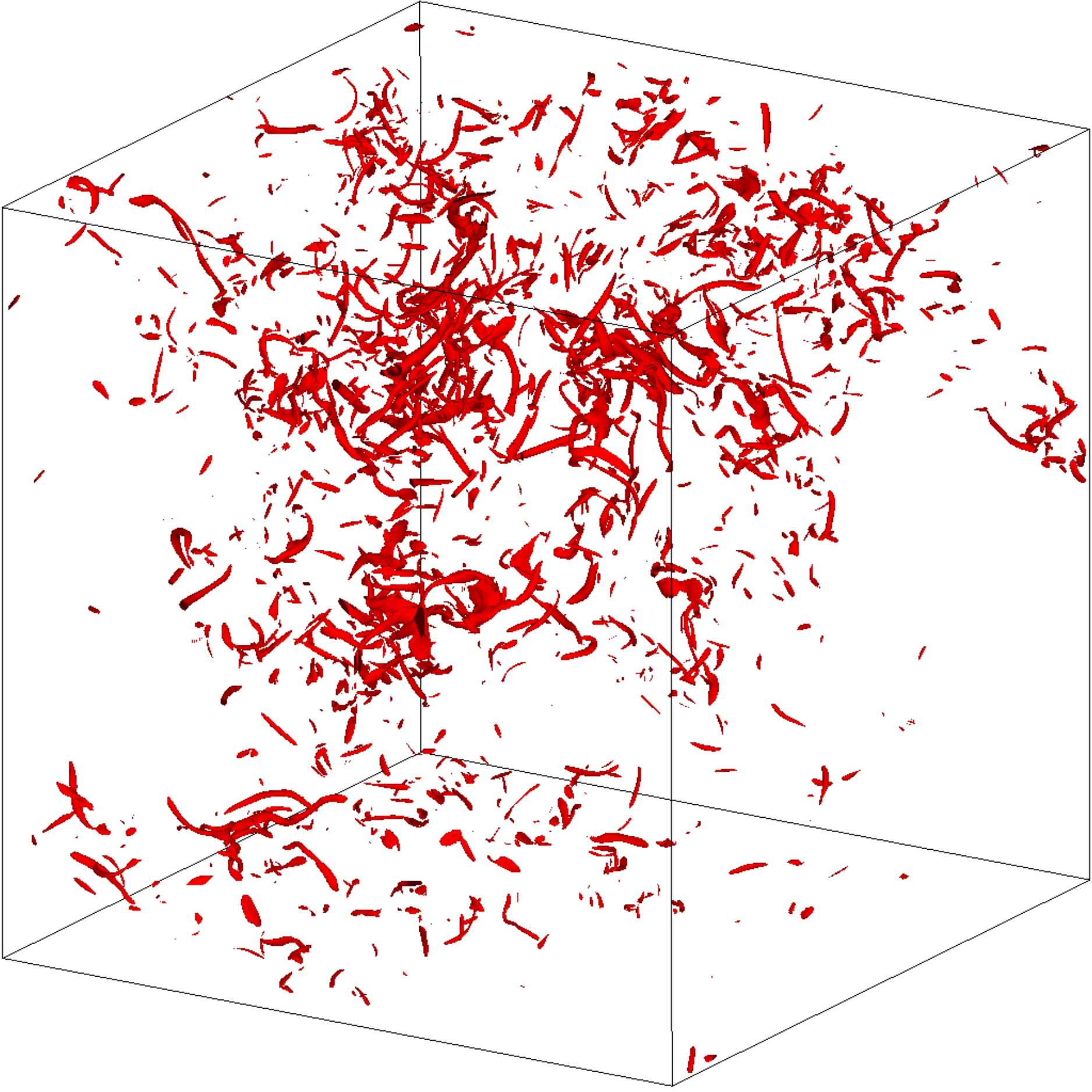}
    \hfill
    \includegraphics[width=0.48\textwidth]{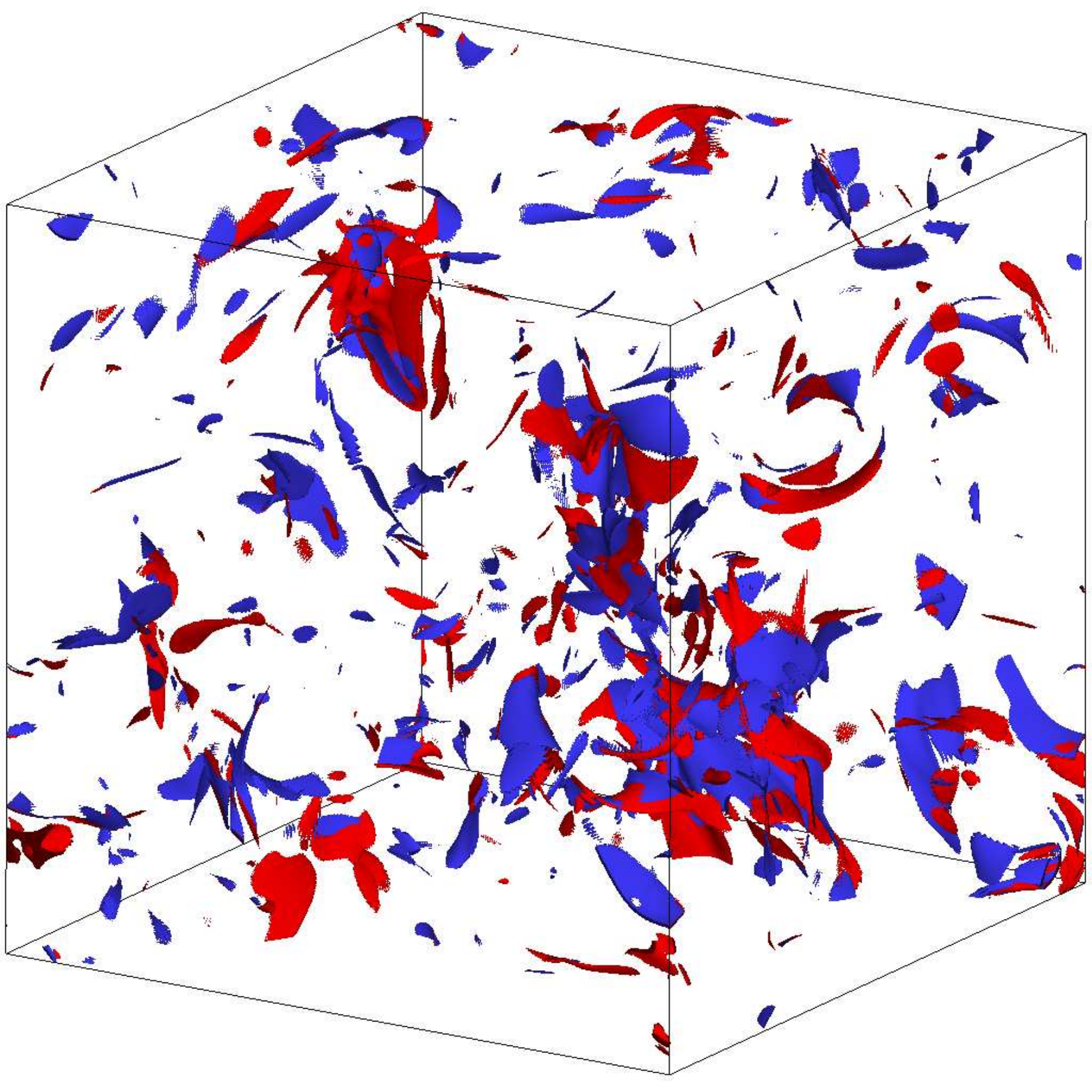}
  \end{center}
  \caption{Nearly singular structures in turbulent flows. Left:
    Vortex filaments in 3D incompressible Navier-Stokes
    turbulence. Shown are iso-surfaces of high vorticity. Right:
    Current sheets in 3D incompressible magneto-hydrodynamical
    turbulence (MHD). Shown are iso-surfaces of high vorticity (red) and
    high current density (blue). The form of the structures and their 
    co-dimension is believed to have major impact on turbulent statistics.
    \label{fig:dissipative-structures}}
\end{figure}

This review is about instantons in fluid flows. Although the physics
of fluid dynamics have little overlap with Yang-Mills theory mentioned
above, they are nevertheless a strongly coupled nonlinear system as
well. The comparison of quantum tunneling in a double well potential
\cite{polyakov:1977} and the Kramer's problem for escape rates in
chemical reactions \cite{kramers:1940,berglund:2013} already
exemplifies the connection between quantum mechanics and stochastic
ordinary differential equations (SDEs). Therefore, the calculus of
instantons in fluid turbulence described by stochastic partial
differential equations corresponds closely to the treatment of
instantons in quantum field theory. However, naturally the question
arises why a non-perturbative method like the instanton approach is
especially promising for tackling the unsolved \textit{problem of
  turbulence}.

In order to justify the approach, we first comment on the statement
that fluid turbulence is one of the most important unsolved problems
in classical physics as stated by R. Feynman. A mathematician, a
physicist and an engineer will all have different thoughts on this
statement. A mathematician will touch the issue by asking questions
like existence of global solutions, bounds on the growth of velocity
derivatives, the limit of vanishing viscosity and minimal regularity
for energy conservation (see e.g.  \cite{bardos-titi:2013,
  eyink-sreenivasan:2006, constantin-e-titi:1994,
  duchon-robert:2000}).  A computational engineer is concerned with
the huge number of degrees of freedom present in a turbulent
flows. Using direct numerical simulations (DNS) only flows with very
low Reynolds numbers can be simulated: a Volkswagen traveling at the
moderate speed of 30 km/h is presently at the absolute limit of direct
numerical simulations. Therefore, constructing and improving
large-eddy simulations using as much insight as possible from the
underlying Navier-Stokes equations is at the heart of computational
modeling \cite{pope:2004}. This understanding is strongly related to
the questions which arise in physics. For a physicist, the turbulence
problem is solved if one has full information on the probability
density of velocity fluctuations or equivalently on correlation
functions of any order. The usual approach known from kinetic theory
is to find some closure approximation for terms involving higher
correlations. In principle, this can be done by applying perturbation
theory starting from the heat equation and introducing the
nonlinearity as perturbation. This approach using renormalized
perturbation theory (direct interaction approximation, DIA
\cite{kraichnan:1964}) and/or renormalization group analysis
\cite{forster-nelson-stephen:1977,yakhot-orszag:1986} was a major
achievement of turbulence theory in the last century.  However, it
seems to be very difficult and questionable, whether these approaches
can describe the probability density functions (PDFs) of velocity
fluctuations in the tail that are extremely different from Gaussian
fluctuations. The reason for this is the occurrence of nearly singular
structures, as depicted in figure \ref{fig:dissipative-structures} for
fluid and plasma turbulence, which are correlated over distances much
larger than the dissipation scale and which cause strong departure
from Gaussian behavior. Thus, for getting access to the tails of the
velocity fluctuations PDF non-perturbative approaches are
required. The instanton approach is especially promising since the
instantons are closely related to nearly singular events like strong
vortex tubes in fluid flows, current sheets in plasmas, and shocks in
high Mach number flows which form the skeleton of anomalous scaling
and departure from Gaussianity, as already noted by Onsager
\cite{onsager:1949,eyink-sreenivasan:2006}.

Before we start to summarize important results using the instanton
calculus e.g. in fluid flows
\cite{daumont-dombre-gilson:2000,shraiman-siggia:1994,
  falkovich-kolokolov-lebedev-etal:1996,gurarie-migdal:1996,
  balkovsky-falkovich-kolokolov-etal:1997, falkovich-lebedev:2011,
  fogedby-ren:2009,wan:2013,bouchet-laurie-zaboronski:2011,
  laurie-bouchet:2015,bouchet-simonnet:2009,
  weeks-tian-urbach-etal:1997, schmeits-dijkstra:2001}, magnetic field
reversals \cite{berhanu-monchaux-fauve-etal:2007}, growth phenomena
\cite{fogedby-ren:2009}, chemical reactions
\cite{miller:1975,kryvohuz:2011}, European options fluctuations
\cite{bonnet-hoek-allison-etal:2005}, and genetic switching
\cite{assaf-roberts-luthey-schulten:2011,gardner-cantor-collins:2000},
we clarify the mathematical background. We start with the path
integral formulation of general stochastic systems pioneered by
Martin-Siggia-Rose/Janssen/de Dominicis. We then identify instantons
as saddle points of the corresponding action and this also naturally
leads to a description known from large deviation theory (LDT) in the
context of the Freidlin-Wentzell theory. After the general procedure
and notations are clarified in section \ref{sec:theory}, we come back
to above mentioned important results in section
\ref{sec:applications}. In \ref{sec:stochastic-burgers}, we will
briefly introduce into a particularly instructive example of
turbulence to show how the instanton formalism is applied to obtain
verifiable predictions. Sections \ref{sec:instanton_computation} to
\ref{sec:higher-D} present techniques to numerically find the
instanton trajectory in fluid dynamical applications by direct
computation, filtering the instanton from numerical experiments, and
in higher dimensional setups, respectively.

\section{Theoretical background}
\label{sec:theory}

\subsection{The Martin-Siggia-Rose/Janssen/de Dominicis formalism}

The functional integral description of stochastic systems started in
the seventies with the Martin-Siggia-Rose/Janssen/de~Dominicis (MSRJD)
formalism \cite{martin-siggia-rose:1973, janssen:1976,
  dominicis:1976}. The method transforms the stochastic system into a
functional integral representation. Although there is little hope to
solve the path integral analytically, it is the starting point for
systematic perturbation theory in stochastic dynamical systems (see
especially \cite{taeuber:2014} for applications in critical dynamics).
It is also the basis for the instanton calculus for stochastic
dynamical systems. To introduce the formalism, consider a stochastic
partial differential equation (SPDE) of the form
\begin{equation}
  \label{eq:spde}
  \dot u = b[u] + \eta(x,t)
\end{equation}
in $d$ space-dimensions, $u$ having $n$ components:
$u(x,t):\mathbb{R}^d \times [-T,0] \rightarrow \mathbb{R}^n$ and $\eta$ being a Gaussian forcing with correlation
\begin{equation}
 \label{eq:correlation}
 \langle \eta(x,t) \eta(x+r,t+s) \rangle = \chi(r)\delta(s)\,,
\end{equation}
i.e. white in time and some correlation function $\chi(r)$ in space.

In fluid dynamics applications, the exact form of the spatial
correlation is often irrelevant and can be characterized solely by its
amplitude $\chi(0)$ and correlation length $L$, where $\chi(r)$ does
not change significantly for $x<L$ and decays to zero quickly for
$x>L$. Dimensionally, for example $L=\sqrt{\chi(0)/\chi''(0)}$
(assuming the forcing is isotropic and $\chi$ depends only on the
scalar distance $r$). The forcing is then completely characterized by
$\chi(0)$ and $\chi''(0)$. Furthermore, we will only be considering
additive noise. The drift $b$ in general is a nonlinear, non-gradient
operator. A (quick) derivation (similar to \cite{ivashkevich:1997}) of
the MSRJD formalism follows by considering the expectation of an
observable $O[u]$ and writing this expectation as a path integral
taken over all noise realizations (using the fact that $\eta$ is
Gaussian and $\delta$-correlated in time). For a more rigorous
discussion of path integrals, in particular in the presented context,
we refer the reader to the works \cite{ticciati:1999,
  albeverio-hoegh-krohn-mazzucchi:2008}. Here, we keep in mind that
the field $u[\eta]$ has a functional dependence on the forcing $\eta$
implicitly given by eqn. (\ref{eq:spde}):
\begin{equation} \label{eq:pathint_mean_}
\langle O[u] \rangle = \int {\mathcal{D}}\eta \, O[u[\eta]] {\mathrm{e}}^{-\int \dotprod{\eta}{\chi^{-1}\eta}/2\, dt}\,,
\end{equation}
where $\dotprod{\cdot}{\cdot}$ is an appropriate inner product,
e.g. $L^2$. At this stage, we can consider the transformation of the
noise paths to the paths of the field $u$ given by $\eta \rightarrow
u$ given by (\ref{eq:spde}), hence $\eta = \dot u - b[u]$. This
coordinate transform leads to a Jacobian in ${\mathcal{D}}\eta = J[u]
{\mathcal{D}} u$ with
\begin{equation}
J[u] = {\mathrm{det}}\left(\frac{\delta \eta}{\delta u}\right)={\mathrm{det}}\left(\partial_t - \frac{\delta b}{\delta u}\right)\,.
\end{equation}
Performing this coordinate change results in the Onsager-Machlup functional 
\cite{onsager-machlup:1953,machlup-onsager:1953}
\begin{equation} \label{eq:pathint_mean}
  \langle O[u] \rangle = \int {\mathcal{D}} u \, O[u] J[u] {\mathrm{e}}^{-\int \dotprod{\dot u - b[u]}{\chi^{-1}(\dot u - b[u])}/2\, dt}\,.
\end{equation}
This formulation is the starting point for directly minimizing the Lagrangian action
\begin{equation}
S_{\cal L}[u,\dot u] = \frac{1}{2} \int \dotprod{\dot u - b[u]}{\chi^{-1}(\dot u - b[u])}\, dt\,.
\end{equation}

Since it is often more convenient to work with the original
correlation function $\chi$ instead of working with its inverse, we
delay this coordinate change and introduce an auxiliary field $\mu$
and obtain (by virtue of the Fourier transform, completion of the square) from
(\ref{eq:pathint_mean}):
\begin{equation} \label{eq:pathint_fourier}
  \langle O[u] \rangle = \int {\mathcal{D}}\eta \, {\mathcal{D}}\mu \,
  O[u[\eta]] \, {\mathrm{e}}^{-\int\left[\dotprod{\mu}{\chi
        \mu}/2-i\dotprod{\mu}{\eta}\right]\,dt}\,.
\end{equation}
Now again considering the coordinate change $\eta \rightarrow u$ we
arrive at the path integral representation of $O[u]$
\begin{equation} \label{eq:pathint_O}
\langle O[u] \rangle = \int {\mathcal{D}}\eta \, {\mathcal{D}}\mu \, O[u] J[u] \, {\mathrm{e}}^{-S[u,\mu]},
\end{equation}
with the action function $S[u,\mu]$ given by
\begin{equation} \label{eq:action_MSR}
S[u,\mu] = \int \left[-i \dotprod{\mu}{\dot u - b[u]} + \frac{1}{2} \dotprod{\mu}{\chi \mu}\right]\,dt\,,
\end{equation}
also termed the \emph{Martin-Siggia-Rose/Janssen/de Dominicis (MSRJD)
  response functional}. In many cases it can be shown explicitly that
the Jacobian $J(u)$ is not relevant to the discussion as it reduces to
a constant, the value of which depending on the choice of either It\=o
or Stratonovich discretization (see e.g. \cite{ivashkevich:1997}).

Two remarks should be added: (i) the change introducing an auxiliary
field $\mu$ to linearize the action with respect to the forcing is
also known as Hubbard-Stratonovich transformation
\cite{stratonovich:1957,hubbard:1959} and (ii) the MSRJD action $S$ is
closely related to classical limit of the Keldysh action
\cite{keldysh:1964,altland-simons:2010,kamenev:2011}.

The main idea of the instanton method is to compute the dominating
contribution to this path integral via a saddle point approximation,
i.e. by finding extremal configurations of the action
functional~(\ref{eq:action_MSR}).

Let us be more specific and consider an observable 
\begin{equation}
  \label{eq:observable-final-time}
  O[u] = \delta(F[u(x,t=0)]-a)
\end{equation}
for a scalar functional $F[u]$, which is defined at the end point
$t=0$ of a path that started at $-T<0$ (keeping in mind that we might
be interested in the limit $T\rightarrow \infty$). The idea is that we
observe an extreme event at $t=0$ that has been created by the noise
(and we give the noise infinite time to create the extreme event). In
turbulent flow, for example, an extreme event of interest could be a
large negative gradient of the velocity profile (i.e.  $F[u] = u_x
\delta(x)$), an event of high vorticity, an event of extreme local
energy dissipation, etc.

Seeking a path integral representation of the probability distribution
${P}(a)$ for the events that fulfill $F[u]=a$ at $t=0$, we
find
\begin{eqnarray*}
{P} (a) &=& \langle \delta(F[u]\delta(t)-a) \rangle \\
&=& \int {\mathcal{D}}\eta \, {\mathcal{D}}\mu \,  \frac{1}{2\pi} \int_{-\infty}^{\infty} d\lambda J[u] \, {\mathrm{e}}^{-S[u,\mu]} \,{\mathrm{e}}^{- i \lambda (F[u]\delta(t)-a)}
\end{eqnarray*}
By now computing functional derivatives, we find 
\begin{eqnarray*}
\frac{\delta S}{\delta \mu} &=& -i \left(\dot u -b[u]\right) + \chi \mu\, \\
\frac{\delta S}{\delta u} &=& i\dot \mu + i(\nabla_u b[u])^{T}\mu 
\end{eqnarray*}
and it is easy to see that the  saddle point equations (the \emph{instanton equations}) are given
by
\begin{eqnarray}
  \label{eq:instanton_u_msr}
        {\dot u} &=& b[u] - i \chi\mu \\ 
        {\dot \mu} &=& -(\nabla_u b[u])^T \mu -i \lambda\nabla_u F[u] \delta(t)\,,\label{eq:instanton_mu}
\end{eqnarray}
where we have incorporated the Lagrange multiplier $\lambda$ at the
right hand side of the equation for $\mu$ as a {\em final}
condition equivalent to
\begin{equation}
{\dot \mu} = -(\nabla_u b[u])^T \mu, \qquad \mu(0) = -i\lambda \nabla_u F[u(t=0,x)]\,.
\end{equation}
A solution $(\tilde u, \tilde \mu)$ of this set of equations is termed
the \emph{instanton configuration} or \emph{instanton}. It represents
an extremal point of the action functional~(\ref{eq:action_MSR}).

We want to make a few remarks on the instanton configuration and the
structure of the chosen path integral representation:

\begin{itemize}
\item In the limit of applicability of the saddle point approximation,
  the instanton configuration corresponds to the most probable
  trajectory connecting the initial conditions to a final
  configuration which complies with the additional constraint defined
  by the observable $O[u]$. In the language of quantum mechanics, it
  corresponds to the \emph{classical trajectory} obtained for the
  limit $\hbar \rightarrow 0$. For stochastic differential equations
  of the form~(\ref{eq:spde}), we similarly need a smallness parameter
  to justify the saddle point approximation. This might either be the
  limit of vanishing forcing, $\chi(0) \rightarrow 0$, or the limit of
  extreme events, $|a| \rightarrow \infty$ (which corresponds to the
  limit $|\lambda|\to \infty$, see discussion in sections
  \ref{ssec:FW} and \ref{ssec:instanton-equations}).
\item It is instructive to apply a change of variables $(u,p) \equiv
  (u,-i\mu)$. The response functional~(\ref{eq:action_MSR}) can then be
  written in terms of a Hamiltonian $H[x,p]$,
  \begin{eqnarray}
    \label{eq:action_MSR_xp}
    S[u,p] &=& \int\left(\dotprod{p}{\dot u} - H[u,p] \right)\,dt, \\ 
    H[u,p] &=& \dotprod{p}{b[u]} + \frac12 \dotprod{p}{\chi p}.
  \end{eqnarray}
  The field variable $u$ and the auxiliary field $p$ then play the
  role of generalized coordinate and momentum for the Hamiltonian
  system defined by $H[u,p]$. The instanton
  equations~(\ref{eq:instanton_u_msr}), (\ref{eq:instanton_mu}) are
  the corresponding Hamiltonian equations of motion. We remark in
  particular that the Hamiltonian $H[u,p]$ is a conserved quantity
  even if the original dynamical system~(\ref{eq:spde}) is
  dissipative.
\item In the above setup, note that the choice $p=0$, $\dot u=b[u]$ is
  a solution of the equations of motion with vanishing action,
  corresponding to a deterministic motion of the original dynamical
  system without any perturbation by noise. Since the action in
  general is always positive, $S[u,p]\ge0$, this implies that a
  deterministic trajectory connecting the initial and final point (if
  it exists) will always be the global minimizer of the action
  functional.
\item Another special solution with $H=0$ is the choice
  $p=-2\chi^{-1}b[u]$, which implies $\dot u=-b$, i.e. the minimizer
  follows reversed deterministic trajectories. This choice is only
  consistent with the auxiliary equation of motion if $\nabla_u b[u] =
  (\nabla_u b[u])^T$, as is easily verified by direct comparison between
  $\dot p$ and equation~(\ref{eq:instanton_mu}). This restriction is
  only fulfilled, if the drift term $b[u]$ is gradient, $b[u] =
  \nabla_u V[u]$, which forms an important sub-class of
  problems. Here, minimum action paths become minimum energy paths.
\end{itemize}

\subsection{Freidlin-Wentzell theory of large deviations}
\label{ssec:FW}

An alternative approach is given by Wentzell and Freidlin
\cite{ventsel-freidlin:1970,freidlin-wentzell:1998}. It has its roots
in the application of large deviation theory \cite{cramer:1938,
  varadhan:2008, dembo-zeitouni:2010} to dynamical systems under
random perturbations. In general, we say that a family of random
processes $u^\epsilon(t)$ defined on $t\in[-T,0]$ fulfills a large
deviation principle, if
\begin{equation}
  \label{eq:ldp}
  P[u^\epsilon(0) \in \Omega] \asymp \exp\left( -\frac1\epsilon \inf_\psi I_T(\psi) \right),
\end{equation}
where ``$\asymp$'' is to be understood as the ratio of the logarithms
of both sides tending to unity as $\epsilon \rightarrow 0$. The left
hand side describes the probability of the random process to ``end
up'' in some set $\Omega$. On the right hand side, the infimum is
taken over all realizations of the path that fulfill the boundary
conditions. The functional $I_T$ is called the \textit{rate function},
which plays the same role as the action functional in the previous
section. The minimizer $\psi_*(t)$ of the rate function represents the
maximum likelihood realization of the random process fulfilling the
boundary conditions, and corresponds to the instanton configuration of
the MSRJD-formalism. In the context of the Freidlin-Wentzell theory,
the rate function can be written down explicitly as
\begin{equation}
  I_T(\psi) = 
  \cases{
    \int_{-T}^0 L(\psi, \dot\psi) dt & if the integral exists\\
    \infty & otherwise,\\
  }
\end{equation}
where
\begin{equation}
  L(x,y) = \sup_p (\dotprod{y}{p} - H(x,p))
\end{equation}
is the Lagrangian of the system with Hamiltonian $H(x,p)$, and
$\dotprod{\cdot}{\cdot}$ is the inner product of the considered space.

We are interested in systems of the form (\ref{eq:spde}),
\begin{equation} \label{eq:sde_wentzel}
  du^\epsilon(t) = b[u^\epsilon(t)]\,dt + \sigma \sqrt{\epsilon} dW\,,
\end{equation}
with a $d$-dimensional Brownian increment $dW$ and $\sigma$ such that
$\chi = \sigma^T\sigma$, where we can write down the Hamiltonian
\begin{equation}
  H(u,p) = \dotprod{b[u]}{p} + \frac12 \dotprod{p}{\chi p}
\end{equation}
and consequently the Lagrangian
\begin{equation}
  L(u, \dot u) = \frac12\dotprod{\dot u-b[u]}{\chi^{-1}(\dot u -b[u])}\,.
\end{equation}
The rate function thus looks like
\begin{equation}
  \label{eq:action_FW}
  I_T(u) = \frac12\int_{-T}^0 \dotprod{\dot u-b[u]}{\chi^{-1}(\dot u     -b[u])}\,dt\,.
\end{equation}

In a similar manner as before we consider observables at the end point
$t=0$ for a scalar functional $F[u^\epsilon(t=0)]$ and want to compute
the probability $P(a)$ of the random process evolving such
that at the final time $F[u^\epsilon(t=0)]=a^*$, with $a^*\in
\mathbb{R}$. As we will see, this corresponds to considering the large
deviations of the moment-generating functions of the random process,
$\langle \exp(\frac\lambda\epsilon F[u^\epsilon(0)]\rangle$, with
$\lambda \in\mathbb{R}$. Starting from the large deviation principle
(\ref{eq:ldp}) we define the set $\Omega_a = \{ \phi | F(\phi) = a^*\}$
of final conditions that comply with the requirement of our
observable. Now, defining $S^*(\lambda) = \epsilon \log \langle
\exp(\frac\lambda\epsilon F[u^\epsilon(0)]\rangle$, for large
$\lambda$
\begin{equation}
  \label{eq:S*}
  S^*(\lambda) = \log \int_\mathbb{R} \exp(\lambda a/\epsilon) P(a) da \sim \max_{a\in\mathbb{R}} (\lambda a/\epsilon - S(a)/\epsilon)\,,
\end{equation}
where $S(a) := -\epsilon \log P(a) = -\epsilon \log
P(u^\epsilon(0) \in \Omega_a)$. The last step of
\eqref{eq:S*} makes use of Laplace's method to show that
$S^*(\lambda)$ is the Legendre transform of $S(a)$, and we conclude
that $\epsilon S^*(\lambda) = \lambda a^* - S(a^*)$. On the other hand
\begin{equation}
  I_T(\psi^a_*) = \inf_\psi I_T(\psi) \sim \frac{\lambda}{\epsilon}a^* - \frac{1}{\epsilon}
  S^*(\lambda) \sim S(a^*)\,,
\end{equation}
where the infimum is taken over all realizations that fulfill the
boundary conditions, $\psi^a_*$ is its minimizer, and of course
$F[\psi^a_*]=a^*$. In short, the rate function corresponds to the
Legendre transform of the logarithm of moment-generating functions. We
obtain the PDF of our observable from the Freidlin-Wentzell action.

To obtain the minimizer $\psi^a_*$, find the solution to the equations
of motion (\emph{instanton equations}) corresponding to
\eqref{eq:action_FW},
\begin{eqnarray}
  {\dot u} &=& b[u] + \chi p \nonumber\\ 
  {\dot p} &=& -(\nabla_u b[u])^T p + \lambda\nabla_u F[u] \delta(t)  \label{eq:instanton_t}\,,
\end{eqnarray}
where the Lagrange multiplier $\lambda$ assures that the constraint
$F[u]=a$ at $t=0$ is satisfied.  We immediately see that these
equations are equivalent to
(\ref{eq:instanton_u_msr}), (\ref{eq:instanton_mu}) by setting $\mu =
-ip$.

\section{Applications of the Instanton method to fluid mechanics and related statistical systems}
\label{sec:applications}

The instanton approach to fluid turbulence started about 20 years ago
with the paper by Gurarie and Migdal \cite{gurarie-migdal:1996}
deriving the scaling of the right tail of the velocity increment PDF
for the Burgers equation \cite{burgers:1974}. For the right tail
shocks are absent and viscosity can be neglected. Even though the same
result was already obtained by other techniques before
\cite{polyakov:1995, feigelman:1980}, this paper contained all
ingredients of the instanton analysis, demonstrating its applicability
in the context of fluid dynamics. A remarkable feature, both of the
right tail of velocity increments and gradients, is that fluctuations
do not play a role and thus that the instanton solution is basically
exact. This is not the case for the left tail of the increment and
gradient PDF. The instanton solution for the left tail dominated by
shocks was introduced in
\cite{balkovsky-falkovich-kolokolov-etal:1997}. The instanton
prediction was possible due to the conservation property of the
Hamiltonian for the instanton equations and the existence of a
Cole-Hopf transformation \cite{cole:1951,hopf:1950}, the latter of
which is in general not available in other fluid models. In the
context of this paper, instantons for the Burgers equation will play a
prominent role, and we will demonstrate many of the discussed methods
and tools in application to Burgers turbulence. The reasons for this
are manifold: As outlined above, many analytical estimates of
instanton results are available for the Burgers equation because of
the mathematical simplicity of the model. Furthermore, not only is the
phenomenology of Burgers turbulence well understood
\cite{bec-khanin:2007} and many open problems of Navier-Stokes
turbulence are known exactly in the Burgers case, but also the
numerical treatment of a 1D model is far less demanding.

Besides the case of Burgers turbulence another problem was studied
intensively during that time: the advection of a passive scalar by a
turbulent velocity field. In a seminal paper, Shraiman and
Siggia\cite{shraiman-siggia:1994} investigated a Lagrangian treatment
of the passive scalar using a path integral approach in the
semi-classical approximation, demonstrating that the passive field as
well as its gradient possess exponential tails. The instanton analysis
was carried out in \cite{falkovich-kolokolov-lebedev-etal:1996,chertkov:1997}
reproducing the results of \cite{shraiman-siggia:1994} but stressing
the relevance of zero modes.

In a similar spirit, there is a corpus of literature applying
instanton calculus to shell models of turbulence. In
\cite{daumont-dombre-gilson:2000} the Gledzer-Ohkitani-Yamada (GOY)
shell model was studied and an iterative procedure was introduced to
determine the instanton.  In addition, quadratic fluctuations around
the instanton were calculated. One interesting outcome of this study
is the tendency towards log-normal statistics of coherent
structures. Similarly, in \cite{biferale-daumont-dombre-etal:1999} a
shell model of passive scalar advection is studied in the context of
instanton calculus. A remarkable and underestimated feature of this
article is the observation that in this case fluctuations around the
instanton have a major impact and dramatically improve the quality of
predictions for the scaling of structure functions.

Instantons for vorticity have been calculated in the inverse cascade
for two-dimensional turbulence \cite{falkovich-lebedev:2011}. We
remark that these instantons are very different from the instantons
considered so far: First, the direct instanton equations for vorticity
are of no use if one considers forcing obtained by taking the
vorticity as observable. The solutions are axial symmetric and the
nonlinearity vanishes, which results in purely Gaussian vorticity
statistics. Therefore, the authors took into account perturbations of
the axial symmetry via the first harmonics and derived an effective
action for which the axial symmetric solution provides an exponential
tail for vorticity statistics. Second, whereas the instanton for
e.g. Burgers turbulence has the property that the action $S$ changes
very rapidly at time $t=0$ this is not the case for the 2D vorticity
instanton. This is a peculiarity of turbulent fluctuations in the
inverse cascade, and thus is neither present in Burgers turbulence,
nor to be expected for the 3D Navier-Stokes instanton.

Instantons have also been calculated for bi-stability and transition
probabilities in geophysical flows \cite{bouchet-simonnet:2009,
  bouchet-laurie-zaboronski:2011, laurie-bouchet:2015}. These
investigations are of enormous importance for blocking phenomena and
multi-state phenomena in atmospheric and ocean flows
\cite{weeks-tian-urbach-etal:1997, schmeits-dijkstra:2001}. These
cases differ from the studies mentioned before, since the instanton is
found by directly minimizing the Onsager-Machlup functional.  A
numerical scheme for this minimum action method in the context of
shear flows was introduced in \cite{wan:2011}, where the author studied
the transition from Poiseuille flow to a traveling wave.  Similar
phenomena, like magnetic field reversals in dynamo action
\cite{glatzmaier-roberts:1995, berhanu-monchaux-fauve-etal:2007,
  benzi-pinton:2011, petrelis-fauve:2010}, wait for an instanton
analysis.  The related problem of growing height profiles described by
the Kardar-Parisi-Zhang equation has been studied in
\cite{fogedby-ren:2009}. Here, the authors applied a numerical
minimization to find instanton solutions and could identify instantons
as nucleations and propagating steps in the height profiles.
Instanton-like or large deviation-like solutions are also intensively
studied in the context of freak waves \cite{mori-janssen:2006,
  mori-onorato-janssen-etal:2007, buehler:2007, peregrine:1983,
  zakharov-gelash:2013, akhmediev-soto-crespo-ankiewicz:2009},
i.e. events of extreme instantaneous ocean surface elevation.

In a broader context, instantons have been extensively studied in
condensed matter and critical phenomena
\cite{leggett:1984,altland-simons:2010,efetov:1999}.  Prototype
examples are the nonlinear sigma model
\cite{weidenmueller-zirnbauer:1988,efetov:1999}, strongly correlated
electrons \cite{mukhin:2008}, and Mott-conductivity
\cite{hayn-john:1991}. Especially, in \cite{hayn-john:1991} the
integrability of the instanton equation allowed also the calculation
of quasi-zero modes and fluctuations. This is in strong contrast to
the situation in turbulence where any analytic solution of the
instanton equations is out of reach. However, this paper motivates to
search for simplified turbulence models such that the corresponding
saddle-point equations are integrable.

The instanton method in the language of large deviation theory with a
focus on a rigorous mathematical treatment has been extensively
studied in the context of statistical mechanics (see especially
\cite{ellis:1985,touchette:2009}). In \cite{touchette:2009} the
relation between the large deviation principle, the saddle point
approximation and Laplace's approximation is discussed.

The calculation of chemical reaction rates based on an instanton
approximation was introduced by Miller \cite{miller:1975}. A recent
successful treatment for multi-dimensional systems is presented in
\cite{kryvohuz:2011}. Furthermore, a field having much in common with
the calculation of chemical reaction rates is genetic switching
\cite{gardner-cantor-collins:2000}. In
\cite{assaf-roberts-luthey-schulten:2011}, an instanton calculation of
switching scenarios has been presented .

We close our discussion with the remark that path-integrals techniques
have found applications in a variety of related fields. These include
financial mathematics (e.g. \cite{bouchaud-potters:2003,
  kleinert:2009, boyle-feng-tian:2007, pham:2007,
  friz-gatheral-gulisashvili-etal:2015}), in particular in relation to
option pricing \cite{linetsky:1997,bonnet-hoek-allison-etal:2005}, but
also other applications in statistical mechanics, for instance
regarding dominant reaction pathways
\cite{autieri-faccioli-sega-etal:2009}, kinetically constrained models
\cite{garrahan-jack-lecomte-etal:2007} (based on the Doi-Peliti
formalism \cite{doi:1976,peliti:1986}), Ginzburg-Landau models
\cite{e-ren-vanden-eijnden:2003, e-ren-vanden-eijnden:2004}, population extinction 
\cite{ovaskainen-meerson:2010}, and nonlinear optics 
\cite{falkovich-etal:2001,falkovich-kolokolov-lebedev-etal:2004,
chertkov-etal:2004,moore:2014, terekhov-etal:2014}.

We finish this section by mentioning that experimental verification
and detection of optimal paths or instantons are still in its infancy.
An exception is the study of a semiconductor laser with optical
feedback \cite{hales-zhukov-roy-etal:2000} where escapes from a
metastable state could be attributed to instantons. We hope that this review 
also motivates and triggers studies in turbulence experiments.

\section{The stochastic Burgers equation}
\label{sec:stochastic-burgers}

As a mathematically less complicated turbulence model we consider the
stochastically driven Burgers equation \cite{burgers:1974} and we will
discuss the presented methods and numerical tools in many cases
applied to Burgers turbulence. In order to understand both the
relevance and the limitations of this model, we will in the following
briefly introduce into the phenomenology of Burgers turbulence.

The stochastically forced, viscous Burgers equation is given by
\begin{equation} 
  \label{eq:stochastic_burgers}
  u_t+uu_x-\nu u_{xx} = \phi\,,
\end{equation}
with a noise field $\phi$ that is $\delta$-correlated in time and has
finite correlation in space with correlation length $L$, more
precisely
\begin{eqnarray} 
  \langle\phi(x,t)\phi(x',t')\rangle &=& \delta(t-t')\chi((x-x')/L), \label{eq:corr_function}\\
  \chi(x) &=& (1-x^2){\mathrm{e}}^{-x^2/2}\,\,. \label{eq:chi}
\end{eqnarray}
While the precise form of $\chi$ is not important, the choice
(\ref{eq:chi}) is common in the literature
\cite{chernykh-stepanov:2001, grafke-grauer-schaefer-etal:2015} and is
used in all numerical computations presented here.

There are several motivations to study Burgers equation. From a
practical point of view, there is a broad range of applications
governed by this model (see e.g. \cite{bec-khanin:2007} for an
overview), for instance in the context of compressible gas dynamics,
in particular weak acoustic perturbations (viewed in the reference
frame of the sound velocity), as well as cosmology
\cite{zeldovich:1970, gurbatov-saichev:1984}, interface growth
\cite{kardar-parisi-zhang:1986} or vehicle traffic flow
\cite{chowdhury-santen-schadschneider:2000}. A major motivation in the
context of fluid dynamics is the fact that Burgers equation has a
similar mathematical structure as the Navier-Stokes equations. As a
one-dimensional equation, however, it is much easier to handle from
both an analytical and computational point of view, in particular
because it contains no non-local pressure term. Therefore, Burgers
equation is often referred to as a {\em testbed} or {\em toy model}
for turbulence.

There are important similarities (and equally important differences)
between the phenomenology of turbulence in fluids described by the
Navier-Stokes equations and Burgers turbulence (see
e.g. \cite{bec-khanin:2007} for an overview of Burgers turbulence
phenomenology). A major similarity between Burgers and the
three-dimensional Navier-Stokes equations is the presence of a direct
energy cascade: If energy is injected on large scales (or low Fourier
modes), this energy is transported via the nonlinearity to small
scales (corresponding to high Fourier modes) until the diffusive part
of the equation becomes important and the energy is
dissipated. Note that the above chosen correlation function
(\ref{eq:corr_function}) models this behavior: In Fourier space we
have $\hat \chi (\omega) = \omega^2 {\mathrm{e}}^{-\omega^2/2}$ such
that we do not have any forcing of the mean (corresponding to
$\omega=0$) and strong decay of the forcing for high frequencies. The
length-scale $L$ of the forcing is dimensionally given by
$L=\sqrt{-\chi(0)/\chi''(0)}$, which in the case (\ref{eq:chi})
evaluates to $L=1$.

Another major similarity between Burgers and Navier-Stokes turbulence
is the existence of intermittency. This is manifest most strikingly in
the existence of anomalous scaling for moments of velocity
differences: The average increment of the velocity field on scale
$h$, $\delta u(h) = \langle u(x+h/2)-u(x-h/2) \rangle$ and its moments
$\delta u_n(h) = \langle (u(x+h/2)-u(x-h/2))^n \rangle$ (``structure
functions'') exhibit a scaling of $\delta u_n(h) \sim h^{\zeta_n}$. In
contrast to the dimensional estimate $\zeta_n = n/3$, the scaling
exponents grow more slowly for $n \rightarrow \infty$ for both Burgers
and Navier-Stokes turbulence, a phenomenon that is believed to be
connected to the intermittent nature of the flow.

In connection to this, a main quantity of interest in fluid systems
are velocity gradients. High gradients are usually related to the most
dissipative structures in the flow which govern the tails of the
underlying probability distributions and structure functions. In the
stochastically driven Burgers equation, we can immediately identify a
difference in the behavior of positive and negative velocity
gradients. For small viscosity $\nu$ and moderate gradients, the
solution will follow the characteristics of the equation, given by the
nonlinearity $uu_x$. This means that positive gradients will be
smoothed out, whereas negative gradients will further steepen until
they are so steep that the viscous term will start to counteract the
advection and shocks are forming. This has important consequences for
the tails of the velocity gradient probability distribution. Let us
fix a point in space-time (for simplicity take $x=0$ and $t=0$) and
ask for the probability to observe a velocity gradient given by $P(a)
= P(u_x(0,0)=a)$.  We are interested in extreme values for $a$, either
positive or negative. Consider first the case of $a>0$. Then, the
noise has to counteract the deterministic dynamics that drive the
system back to smaller positive gradients. Intuitively, we will find
that it is very difficult for the noise to generate such gradients,
and we may expect the probability density to decay faster than a
Gaussian. For sufficiently large $a$, the scaling of the tail of the
probability distribution should be characterized by the scaling of the
associated minimizer of the Freidlin-Wentzell functional as we are
clearly in a regime where we expect a large deviation principle to
apply. The left tail of the velocity distribution is expected to have
two different regimes: For small viscosity, it should be relatively
easy for the system to generate moderate negative velocity gradients,
simply following the deterministic dynamics of the nonlinearity that
steepens the profile of the solution which would eventually lead to
discontinuities in the velocity field if the system did not have any
viscosity. Since the viscosity prohibits the occurrence of infinite
gradients, once the gradient becomes too steep, it is again difficult
for the noise to produce large negative gradients. Then, in the {\em
  viscous tail} of the distribution, again, the large deviation
principle should be applicable and the scaling behavior is expected to
be captured by the instanton (or minimizer of the Freidlin-Wentzell
functional).

Following this intuition, there has been a considerable body of work
devoted to the detail of the scaling of the function $P(a)$. Here, we
focus on several important studies that had consequences in the
context of the present review. Concerning the right tail scaling,
Gurarie and Migdal \cite{gurarie-migdal:1996} used the MSRJD formalism
in order to derive the Euler-Lagrange equations associated with
Burgers equation (\ref{eq:stochastic_burgers}) which are given by
\begin{equation}
  u_t +uu_x-\nu u_{xx} = \chi p, \qquad p_t +up_x+p_{xx} = -\lambda
  \delta'(x) \delta(t) \,.
\end{equation}
These equations follow directly from (\ref{eq:instanton_t}): simply
set $b[u] = -uu_x+\nu u_{xx}$ and compute $(\nabla_u b)^T =
u\partial_x + \nu\partial_{xx}$ using integration by parts. In their
work, focusing on the right tail of the probability distribution $p$,
Gurarie and Migdal introduced a finite-dimensional dynamical system
approximating the solution of the Euler-Lagrange equation in order to
predict that the distribution should decay much faster than Gaussian,
i.e.
\begin{equation}
  \ln(P(a)) \sim -(a/\omega)^3, \qquad \omega^3 = -\frac12 \chi''(0)\,.
\end{equation}
For the viscous left tail, Balkovsky et
al. \cite{balkovsky-falkovich-kolokolov-etal:1997} applied the
Cole-Hopf transform to the instanton equations and used a variety of
careful approximations in order to predict that, in the limit
$a\rightarrow -\infty$, the scaling of $p$ should behave like
\begin{equation}
  \ln(P(a)) \sim -(a/(\omega {\mathrm{Re}}))^{3/2}, \qquad
     {\mathrm{Re}} = L^2\omega/\nu\,.
\end{equation}
These limiting results can be motivated by a rather simple
phenomenological description
\cite{balkovsky-falkovich-kolokolov-etal:1997}: Velocity differences
$\delta u(h) = u(h/2)-u(-h/2)$ at the length scale $h$ are increased by
the (Gaussian) forcing according to the law $\delta u^2 \sim \phi^2
t$. The breaking time of the shock structures can be estimated by $t
\sim L/\delta u$. Therefore, from $P(\phi) \sim \exp(-\phi^2/\chi(0))$,
one obtains $P(\delta u) = \exp(-\delta u^3/(L \chi(0))$. Now, in
smooth regions $u_x \sim \delta u/h$, while for the shocks $u_x \sim
-\delta u^2/\nu$. We thus recover the exponents $3$ and $3/2$ for the
right and left tail, respectively.

Yet, when comparing these limiting results to measurements in DNS,
until recently, the role of the instanton for negative velocity
gradients was unclear (and actually actively discussed among
researchers). One numerical result obtained by Gotoh \cite{gotoh:1999} via
massive direct numerical simulations presented an estimate of the
local scaling exponent of $1.15$ for the probability distribution of
the negative velocity gradients, which is surprisingly far away from
the analytical prediction of $3/2$. In 2001, Chernykh and Stepanov
developed a numerical scheme to solve the instanton equations
numerically \cite{chernykh-stepanov:2001}. This way they were able
to show that all the approximations made by Balkovsky et al. leading
to a scaling exponent of $3/2$ were valid for the solution of these
equations. These results rendered the discrepancy between DNS
measurements and theoretical prediction even more in need of an
explanation: In what sense are instanton configurations actually
relevant in Burgers turbulence? In our recent work, we were able to
confirm the relevance of instantons and their impact on the tails of
the velocity gradient distribution in two ways: First, we showed that
the instanton configurations themselves can be identified in
realizations of turbulent Burgers flows by using an appropriate
filtering technique \cite{grafke-grauer-schaefer:2013}. Second we were
able to analyze Gotoh's simulation in detail and compare them to the
local scaling exponent given by the solution of the instanton
equations \cite{grafke-grauer-schaefer-etal:2015}. It turned out that,
for the parameters that were chosen in Gotoh's simulation, the scaling
exponent of the velocity gradient given by the (numerical) solution of
the instanton equations is actually $1.16$, hence very close to the
measured value. The regime in which these numerical simulations were
carried out was simply not yet in the range of validity of the
asymptotic analysis that was carried out analytically, but
nevertheless already in a regime where the instanton approximation is
valid. The resolution of this 'puzzle' is encouraging and gives hope
that instantons are relevant in a wide-range of fluid dynamics and can
help to answer many open questions in the field. However, the detailed
study of the stochastic Burgers equation also clearly showed how
important it is to develop efficient numerical methods in order to
compute such instantons.

The identical question was discussed for the tail scaling behavior in
the inviscid limit, $\nu \rightarrow 0$. Phenomenologically, the
exponential tail decay is only valid for viscosity dominated shocks
and the exponent of $3/2$ appears due to the interplay of shock
steepening by the non-linearity and shock smoothing by the dissipative
term. In the inviscid limit, the exponential region of the PDF is
pushed to negative infinity, and for finite gradients an algebraic
tail $P(a)\sim a^{-\gamma}$ prevails. There was a debate about the
numerical value of $\gamma$, with predictions ranging from $\gamma=2$
\cite{bouchaud-mezard:1996, yakhot-chekhlov:1996}, over $\gamma
\approx 3$ \cite{polyakov:1995, boldyrev:1997, gotoh-kraichnan:1998}
to $\gamma=7/2$ \cite{e-khanin-mazel-etal:1997}. The issue was settled
in \cite{e-vanden-eijnden:2000} by considering the scaling of
pre-shocks and confirmed numerically
\cite{bec-frisch-khanin:2000}. Notably, field-theoretic methods
\cite{polyakov:1995} have resulted in wrong predictions and it is
unclear in what manner the instanton formalism can be applied to
obtain the correct exponent in this limit.

In the next section we will first present recent developments of
efficient numerical techniques for solving the instanton
equations. After that, we will discuss how to obtain instantons by
filtering from direct numerical simulations and we will show how these
two methods can be compared. The remainder of the paper is dedicated
to higher-dimensional problems which occur naturally when considering
many models of physical fluids.

\section{Efficient numerical solution of the instanton equations}
\label{sec:instanton_computation}

The problem of the numerical solution of the instanton equations, or,
equivalently, the numerical minimization of the action functional,
lies at the very center of the computation of instanton
configurations. In the last decades, a multitude of numerical schemes
were proposed, differing both in the setup and approach as well as in
practical considerations, such as computational efficiency or
parallelizability. In this section, we want to briefly introduce into
this spectrum of methods, starting from traditional and general
methods such as shooting methods \cite{keller:1992} to very recent
developments \cite{vanden-eijnden-heymann:2008,
  heymann-vanden-eijnden:2008-b,
  grafke-grauer-schaefer-etal:2014}. Differences in terms of
applicability, scope and practicability are highlighted in order to
allow a comparison of different schemes for a given problem in terms
efficiency and simplicity. Furthermore, we particularly highlight
specific techniques for efficiently computing the instanton field
configuration for fluid-dynamical problems, giving practical
tips. These include implementation details for reoccuring problems
such as computing norms in function spaces for degenerate forcing,
treatment of numerically unfavorable correlation matrices and
increasing computational and memory efficiency. This discussion will
be complemented with an example implementation given in the appendix.

Next to the possibility of computing rare and extreme events from the
knowledge of the action functional or the corresponding equations of
motion, alternatives exist that do not rely on the saddle-point
approximation of the underlying path-integral. One notable class of
algorithms approximates the infinite-dimensional path-integral
\eqref{eq:pathint_O} numerically, for example by Monte-Carlo
integration. Algorithms of this type are regularly used for
calculations in quantum chromodynamics (``lattice QCD''), but have
recently been applied to fluid-dynamical problems as well
\cite{mesterhazy-jansen:2011}. Another class of algorithms devises
strategies to increase the number of rare events by considering an
ensemble of trajectories, and dynamically cloning and destroying
realizations according to an empirical estimator of the rare event
\cite{moral-garnier:2005, giardina-kurchan-lecomte-etal:2011,
  simonnet:2014, rolland-simonnet:2015}. Suited also for the
application in fluid dynamics, they can be seen as the counterpart of
the instanton formalism in estimating the far tails of probability
densities. As both these classes of methods in general do not make use
of the instanton approximation, we will not discuss them in more
detail here, even though many of them are applicable for the
computation of rare and extreme events in fluid dynamics.

\subsection{Gradient systems, minimum energy paths}

In the context of theoretical chemistry and the computation of
reaction rates, a common scenario is a simplification of
system~(\ref{eq:spde}), where the drift-term is a gradient,
$b[u]=-\nabla_u V(u)$. Here, minimum action paths can be translated into
minimum energy paths, which in turn allows for a number of
simplifications of the numerical method, as a lot more is known about
the properties of the minimizer: Between stable fixed points of the
deterministic dynamics, which correspond to the minima of the
potential landscape $V(u)$, the transition trajectory has to cross
through the saddle point $u_*$ of the potential with the lowest
potential value $V(u_*)$, when moving between neighboring basins of
attraction. This simplified structure of the minimizer allows for a
direct estimate of the associated transition probability in terms of
the height of the potential barrier and the Hessian at the saddle
point, ultimately leading to Kramers' law \cite{kramers:1940}, which
is a refinement of Arrhenius' law, known from the problem of chemical
reaction rates \cite{arrhenius:1889}. Even though in the context of
fluid dynamics the systems involved are usually not gradient, some of
the mentioned methods are used as a basis for more generalized
schemes. Notable methods for finding minimizing trajectories in this
setup include the nudged elastic band method
\cite{henkelman-jonsson:2000} and the string method
\cite{e-ren-vanden-eijnden:2002, e-ren-vanden-eijnden:2007}.

\subsection{Numerical computation of the minimizer of the action functional}
\label{ssec:direct_minimization}

Returning to the problem of an arbitrary action functional, more
general methods have been developed, most notably the minimum action
method (MAM) \cite{e-ren-vanden-eijnden:2004}, which can be seen as a
generalization of \cite{olender-elber:1996} and variants, for example
adaptive MAM \cite{zhou-ren-e:2008, wan:2011}, parallel MAM
\cite{wan-lin:2013}, and gentlest ascent dynamics (GAD)
\cite{e-zhou:2011}.  The main idea for an efficient computation is to
exploit the structure of the underlying Euler-Lagrange (instanton)
equations (\ref{eq:instanton_t}) for efficient computation of the
solution. This structure might depend on the particular problem and
thus, the computational method often needs to be adapted to the
concrete problem.  An important step in the development of a method
that can be applied in a very general context is the {\em geometric
  minimum action method} (gMAM). The gMAM can be viewed as a
generalization of the string method for non-gradient fields. Its
starting point is the construction of the quasi-potential that exists
even if the drift $b$ does not possess a potential (for details see
chap. 5 of \cite{freidlin-wentzell:1998}). If the drift $b$ does not
depend explicitly on time, time can be viewed as a parametrization of
the instanton equations. This parametrization can, in principle, be
changed to one that is more favorable from a computational point of
view.  Based on this idea, the gMAM
\cite{vanden-eijnden-heymann:2008,heymann-vanden-eijnden:2008-b,heymann-vanden-eijnden:2008}
was developed which offers an efficient way to compute minimizers for
transitions between an initial and a final state and works even in
cases where $b$ does not possess a potential. In particular, one
arrives at a modified action functional,
\begin{equation}
  \label{eq:gmam_action}
  \hat S[u,\dot u] = \int_0^1 \left( \|\dot u\|_{\chi} \|b[u]\|_{\chi} -
  \dotprod{ \dot u}{b[u]}_{\chi} \right) ds,
\end{equation}
where the search space is restricted to arclength parametrized
trajectories. The metric is given by the correlation matrix $\chi$
of the problem via
\begin{equation}
  \dotprod{u}{v}_{\chi} \equiv \dotprod{u}{\chi^{-1} v)}_{L^2}\,,
\end{equation}
and the norm $\|\cdot\|_{\chi}$ is induced by this scalar product. The
action functional~\eqref{eq:gmam_action} is also termed the
\emph{geometric action functional}, and its minimizers are identical
to the infinite time minimizers of the original action, if
reparametrized to physical time.

An important class of problems that can be solved efficiently by the
gMAM are noise-driven transitions between stable fixed points in the
context of meta-stability. These problems are difficult to solve
computationally in the original parametrization using time, since it
can be shown that it takes infinite time for the minimizer to approach
the stable fixed point. To illustrate this fact, in the simplest case,
consider a one-dimensional Ornstein-Uhlenbeck process $du = -\gamma u
\,dt + \sqrt{\epsilon} dW $ for the exit from a stable fix point
$u=0$, hence the transition from $u(T_{\min})=0$ at $T_{\min}<0$ to
$u(0)=a > 0$. In this case, the explicit solution of the instanton
equations yields
\begin{equation}
  u(t) = a{\mathrm{e}}^{\gamma t}\left(\frac{1-{\mathrm{e}^{2\gamma (T_{\min}-t)}}}{1-\mathrm{e}^{2\gamma T_{\min}}}\right)\,,
\end{equation}
such that the action $S$ depends on $T_{\min}$, hence
\begin{equation}
S(T_{\min}) = \frac{\gamma a^2}{\epsilon(1-{\mathrm{e}}^{2\gamma T_{\min}})}, \qquad \inf_{T_{\min} \in (-\infty,0)} S(T_{\min}) = \frac{\gamma a^2}{\epsilon}\,.
\end{equation}
The geometric reparametrization used in both the string method and the
gMAM avoids the computational difficulties that arise from the fact
that the minimizer requires $T_{\min}\rightarrow \infty$, which
makes them in particular efficient for the problems involving the
transition between fix points.

\subsection{Solution of the instanton equations}
\label{ssec:instanton-equations}

As well known from classical mechanics, the alternative to minimizing
the action functional is the solution of the corresponding equations
of motion, which are equivalent to the instanton equations
\eqref{eq:instanton_t}. This dualism transfers into the algorithmic
context: Instead of employing global numerical minimization of the
action functional, we can choose to solve a pair of coupled
deterministic PDEs instead. In general, it is not obvious which choice
is advantageous. In the context of fluid problems, though, we will
show in the following that many problems are more easily treated by
solving the equations of motion.

At the center of this lies the realization that commonly we are
interested in observables of the form of
\eqref{eq:observable-final-time}, measuring a single degree of freedom
in the final field configuration. Most importantly, we do not want to
specify the full final condition of our field, or have no access to it
\emph{a priori}. For example when finding dissipative structures in
Burgers turbulence, we do not know the exact form of the shock
structure that will evolve. While the creation of a steep gradient can
still be viewed as an exit problem (interpreting the initial state of
the system $u_1=u(T_{\min},x)=0$ as the stable fix point for
$T_{\min}\rightarrow -\infty$), we would choose $O[u] =
\delta(F[u(x,t=0)]-a)$ as observable with $F[u(x,t=0)] = \partial_x
u(x=0,t=0)$, notably without specifying the final state $u_2$. The
choice of the observable restricts $u_2$ to have a gradient equal to
$a$ at $x=0,t=0$, but makes no other assumption about its form. In
other words, while it is intuitively plausible that the system will
develop a diffusion-limited shock-like structure in order to produce a
large negative gradient at $t=0$, we cannot compute $u_2$ without
actually obtaining the full solution of the instanton equations
(\ref{eq:instanton_t}). This is a quite powerful realization: The only
input being a restriction on a single degree of freedom of the final
condition, the instanton formalism (if applicable) provides us with
the most probable final state which fulfills the constraint, the most
probable evolution in time from a given initial condition into this
state, and the force that was necessary to achieve this evolution.

As a direct consequence, problems of this type are difficult to frame
in the context of minimization of the action functional, as the final
condition is not fixed, but only subject to a constraint. In contrast
to this they are readily solved by considering the instanton
equations \eqref{eq:instanton_t}
\begin{eqnarray*}
  {\dot u} &=& b[u] + \chi p\\ 
  {\dot p} &=& -(\nabla_u b[u])^T p + \lambda\nabla_u F[u] \delta(t)
\end{eqnarray*}
as the term $\lambda\nabla_u F[u] \delta(t)$ defines an \emph{initial
  condition} $p(x,t=0)=-\lambda\nabla_u F[u]$ for the auxiliary equation,
propagated \emph{backwards} in time. The problem is therefore reduced
to solving two mutually dependent deterministic PDEs with known
initial conditions. In what follows, we explain in detail how to
implement this formalism in the case of Burgers shocks, noting that
the scheme is similarly applicable to different problems in fluid
dynamics of the same functional form.

In 2001, Chernykh and Stepanov published a seminal paper describing an
algorithm suited for such boundary conditions
\cite{chernykh-stepanov:2001}.  Their idea was to solve
(\ref{eq:instanton_t}) iteratively, demonstrating that the iterative
scheme converges to the instanton given by the solution of
(\ref{eq:instanton_t}). The original motivation of Chernykh and
Stepanov was to confirm analytical predictions regarding the scaling
exponent of $3/2$ of the left tail of the velocity gradient, when
diffusion counteracts the steepening of the shock profile. For the
case of extreme negative velocity gradients, they chose $O[u] =
\delta(F[u(x,t=0)]-a)$ with $F[u(x,t=0)] = \partial_x u(x=0,t=0)$ and
$a \ll 0$. This choice leads to the boundary condition $p(x,t=0) =
-\lambda \delta'(x)$. In order to iteratively solve the equations, the
basic scheme suggested by Chernykh and Stepanov was to view the
instanton equations (\ref{eq:instanton_t}) from a dynamical point of
view on the time interval $(-\infty, 0]$. For a specified velocity
field, for example the field $u^{(k)}(x,t)$ of the $k$-th iteration,
we can solve the auxiliary equation backwards in time up to a time
$T_{\min}\ll 0$. From our previous considerations of the simple
Ornstein-Uhlenbeck case, it is clear that one would like to choose
$T_{\min}$ as small as possible in order to approximate the minimizer
(which is obtained in the limit $T_{\min}\rightarrow -\infty$).  Once
the next iteration $p^{(k+1)}(x,t)$ is found on $[T_{\min}, 0]$, this
solution is used in the equation for $u$, to find the next iterate
$u^{(k+1)}(x,t)$. Note that the convolution $\chi p$ represents the
forcing of the system, while the rest of the $u$-equation resembles
the deterministic part of the original Burgers equation
\eqref{eq:stochastic_burgers}. Therefore, having found the fixed point
of our iteration, we can recover the \emph{optimal forcing} by
evaluating $\chi p$. To start off the iteration, an initial guess
$u^0(x,t)$ is required, e.g. $u^0(x,t)=0$. The iteration is continued
until reaching a fixed point. The following figure illustrates the
iteration scheme:
\begin{center}
\begin{tikzpicture}
  \node (m-1-1) at (-4,1.5) {$u(x,T_{\min})$};
  \node (m-1-2) at (-2,1.5) {};
  \node (m-1-3) at ( 2,1.5) {$u(x,t)$};
  \node (m-2-1) at ( 4,0) {$p_0(x)=-\lambda \delta'(x)$};
  \node (m-2-2) at ( 2,0) {};
  \node (m-2-3) at (-2,0) {$p(x,t)$};
  \path[-stealth] (m-1-1) edge node [above] {$u_t + uu_x -\nu u_{xx} = \chi p$} (m-1-3)
   (m-2-1) edge node [below] {$p_t + up_x  + \nu p_{xx}= 0$} (m-2-3)
   (m-2-3) edge (m-1-2)
   (m-1-3) edge (m-2-2);
\end{tikzpicture}
\end{center}
Note that one needs to specify the value of $\lambda$ in this
algorithm as it is present in the term $p(t=0,x) = -\lambda
\delta'(x)$.  Usually, it is not known {\em a priori} how $\lambda$
relates to the parameter $a$ in the observable $O[u] =
\delta(F[u(x,t=0)]-a)$. One can view $\lambda$ as the Lagrange
multiplier that ensures that at time $t=0$ we are observing indeed
$F[u(x,t=0)] = a$.  In the above scheme, however, one can measure $a$
after each iteration by computing $F[u^{(k)}(x,t=0)]$. Once the
differences in the values of $a$ are below a chosen threshold, the
desired approximation of the instanton has been computed. An example
implementation is given in the appendix in listing \ref{lst:OUiter}
for the case of a non-linear Ornstein-Uhlenbeck process.

Although this scheme appeals by its simplicity and generality, there
are several computational challenges associated with it.  For the
Burgers case, this was already noted by Chernykh and Stepanov. Some of
them are so severe that they make a direct application of the scheme
almost impossible. In the following, we discuss two particular
challenges: (a) the numerical instability of the fixed point and (b)
convergence issues related to the fact that one needs to take
$T_{\min}\rightarrow -\infty$ in order to compute the minimizer.

Regarding (a), we confirmed in our simulations results from Chernykh
and Stepanov: For the Burgers case, for large $\lambda$ (corresponding
to large negative gradients $a$), taking directly the updated
$u^{(k+1)}$ as the new advection field in the equation for $p$ can
lead to instabilities. These instabilities can be suppressed by a
softer update in the spirit of a Gauss-Seidel iteration: Assume that,
in the above scheme, we have computed the new velocity field $u$ on
$(T_{\min}, 0]$ and let us call this field $\tilde u ^{(k+1)}$. Then
  we can take as an update a weighted average of the newly computed
  field $\tilde u ^{(k+1)}$ and the old field $\tilde u ^{(k)}$ using
  a weight parameter $\alpha \in [0,1)$
\begin{equation}
  u^{(k+1)} (x,t) = \alpha u^{(k)}(x,t) + (1-\alpha)\tilde u ^{(k)} (x,t)\,.
\end{equation}
During the iterative procedure, we monitor the numerical value of the
action at each iteration step to ensure that the update yields a
minimizer. Note that one could have also used a gradient based method
to descent into the minimizer. However the choice above enables us to
re-use existing implementations of the deterministic dynamics.

Regarding (b), the difficulties associated with $T_{\min}\rightarrow
-\infty$ require a more extensive modification of the algorithm. One
approach is to borrow an idea from geometric methods, in particular
the string method and the minimum action method (gMAM), in order to
find a suitable reparametrization of the evolution variable
\cite{grafke-grauer-schaefer-etal:2014}. Hamilton's equations, which
describe the trajectory of the minimizer of the Freidlin-Wentzell
functional, are given by
\begin{equation}
  \dot u = \frac{\partial H}{\partial p}, \qquad \dot p = - \frac{\partial H}{\partial u}\,.
\end{equation}
If $H$ does not depend explicitly on time, we can reparametrize these
equations using $t=t(s)$ (assuming $t'(s)>0$) to obtain
\begin{equation} \label{arc-length-parametrized-hamilton}
  u' = t'(s) \frac{\partial H}{\partial p}, \qquad p' = -
  t'(s) \frac{\partial H}{\partial u}\,.
\end{equation}
We know from the geometric minimum action method that a suitable
parametrization to circumvent the computational difficulties
associated with $T_{\min}\rightarrow -\infty$ is to choose $t(s)$ in
such a way that $\|u'\|=C$. The norm $\|\cdot \|$ is simply the
Euclidean length for $\delta$-correlated noise. For a correlation
function $\chi$, the induced norm $\|\cdot\|_{\chi}$ is given by
$\|f\|_{\chi} ^2= \langle f,\chi^{-1} f\rangle$ and we choose $t(s)$
such that $\|u'\|_{\chi}=C$. In analogy, we call this {\em
  arclength-parametrization} of Hamilton's equations for the
minimizer. For the Burgers case, working directly with the correlation
function given in (\ref{eq:corr_function}) can lead to instabilities
for high frequencies. Therefore, one can truncate the correlation
function in Fourier space, e.g. work with
\begin{equation} \label{eq:corr_function_mod}
  \hat\chi(\omega) = \omega^2{\mathrm{e}}^{-\omega^2/2}{\mathcal{H}}(\omega_c-|\omega|)\,,
\end{equation}
where $\mathcal{H}$ denotes the Heaviside function. Then, the
correlation function $\chi$ can be inverted on its support in Fourier
domain and we can implement the induced norm $\|\cdot\|_{\chi}$ via
\begin{equation}
  \|f\|_{\chi} ^2 = \left(\left\langle f, {\mathcal{F}}^{-1}\left(\hat\chi^{-1}\hat f\right)\right\rangle\right)^{1/2}\,,
\end{equation}
where $\hat f$ is the Fourier transform of $f$ and
${\mathcal{F}}^{-1}$ denotes the inverse Fourier transform. Then, for
Burgers equation, the arclength-parametrized Hamilton equations
(\ref{arc-length-parametrized-hamilton}) are written as
\begin{equation}
  \label{eq:instanton-burgers-geometric}
  u_s = \frac{\|u_s\|_{\chi}}{\|b[u]\|_{\chi}}(b[u]+\chi p), \qquad p_s = - \frac{\|u_s\|_{\chi}}{\|b[u]\|_{\chi}} (up_x + \nu p_{xx})\,.
\end{equation}
Note that it follows from the above Hamilton's equations that we have
at the minimizer
\begin{equation}
  \|b[u]\|_{\chi} = \|b[u]+\chi p \|_{\chi}\,.
\end{equation}
This equation can be used in order to regularize the term
$\|b[u]\|_{\chi}$ in numerical simulations if necessary. The arclength
parametrized equations of motion
\eqref{eq:instanton-burgers-geometric} can similarly be obtained by
considering the variation of the geometric action
\eqref{eq:gmam_action}.

It is noteworthy that Chernykh and Stepanov were aware of the
difficulties related to taking $T_{\min}\rightarrow -\infty$. They
addressed this difficulty in their paper in two ways. First, they
replaced the initial condition for $u$ at $T_{\min}\rightarrow
-\infty$ with a modified initial condition found from the
linearization of the instanton equations around the fixed
point. Second, they used a self-similar transform related to the
kernel of the heat equation which resulted in an exponential rescaling
of the time. While this approach proved to lead to efficient numerics
for the Burgers case, the above chosen geometric reformulation leads
to similar results but furthermore generalizes to other equations in a
straightforward way.

We remark that for known initial and final conditions of the field
variable, the method lined out above is not easily applied. In this
case there is no restriction on the auxiliary variable, while the
field variable has to hit a particular point $u_2$ starting from
$u_1$. To solve the instanton equations, one can then only rely on
shooting methods \cite{keller:1992}, which is nearly hopeless in such
high-dimensional systems. Therefore, the direct minimization
techniques discussed in section \ref{ssec:direct_minimization} are in
general preferred.

\section{Instanton filtering in direct numerical simulations}
\label{sec:instanton_filtering}

Instantons describe the rare fluctuations of a system, far from its
equilibrium state. For complicated systems, like fluids, it is
difficult to predict analytically at which point the instantons will
start to govern the behavior of the underlying probability
distributions. One can even raise the question whether, in any
realistic context, they are relevant to describe phenomena as they
might become only important so far out in the tails that, while
presenting a correct mathematical description of the asymptotic
behavior of the tails, they are not of any relevance for understanding
or modeling the physical phenomena. As lined out in section
\ref{sec:stochastic-burgers}, precisely this question was raised and
answered in the context of the stochastic Burgers equation regarding
the left tail of the velocity gradients. From a practical and
numerical point of view, however, there is an entirely different way
to assess the importance of instantons: If one creates a very large
number of direct stochastic realizations of the system, one can,
fairly easily, filter the runs (or paths) that lead to a desired value
of the observable. The ensemble of these numerically filtered -- or
conditioned -- paths represents an approximation of the path integral
under the appropriate boundary conditions and will therefore resemble
the instanton whenever the saddle point approximation applies. It is
clear from the discussion above that, in order to carry out this
scheme without any further optimization, massive numerical simulations
are needed. Stochastic Monte-Carlo simulations, however, parallelize
in a trivial way and, with the rapid development of high-performance
computing over the last decade, direct numerical simulations are now
feasible as we will illustrate using the stochastic Burgers equation.

\subsection{Illustration of numerical filtering}

In order to show the simplicity of the numerical filtering, let us consider a simple one-dimensional Ornstein-Uhlenbeck process, hence
\begin{equation} \label{eq:sde_bm_OU}
du = b(u)\,dt + \sigma dW\,, \qquad b(u) = -\gamma u\,, \qquad \gamma > 0\,.
\end{equation}
We consider the exit from a stable fixed point at $\lim_{t \rightarrow   -\infty} u(t) = 0$ such that $u(0) = a$.  A simple analytic solution of the instanton equations~(\ref{eq:instanton_u_msr}), (\ref{eq:instanton_mu}) shows that the minimizer of the Freidlin-Wentzell action is given by
\begin{equation}
  u(t) = a{\mathrm{e}}^{\gamma t}\,.
\end{equation}
\begin{figure}[tb]
  \centering
  \includegraphics[width=0.48\textwidth]{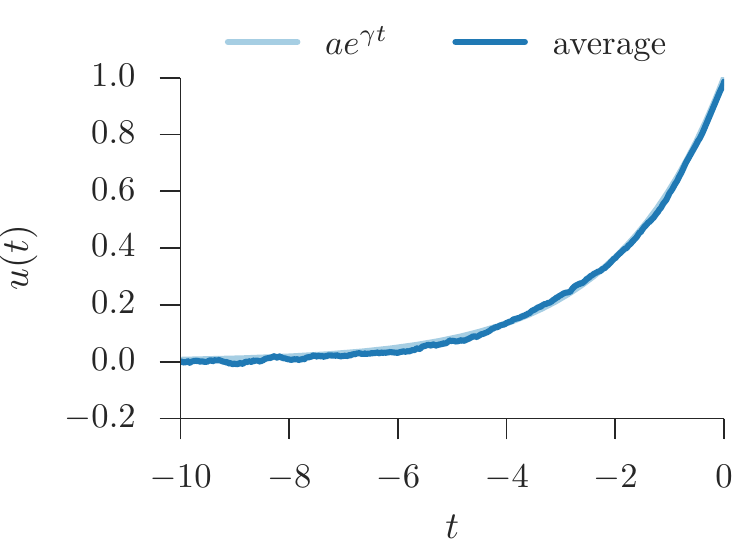} 
  \hfill
  \includegraphics[width=0.48\textwidth]{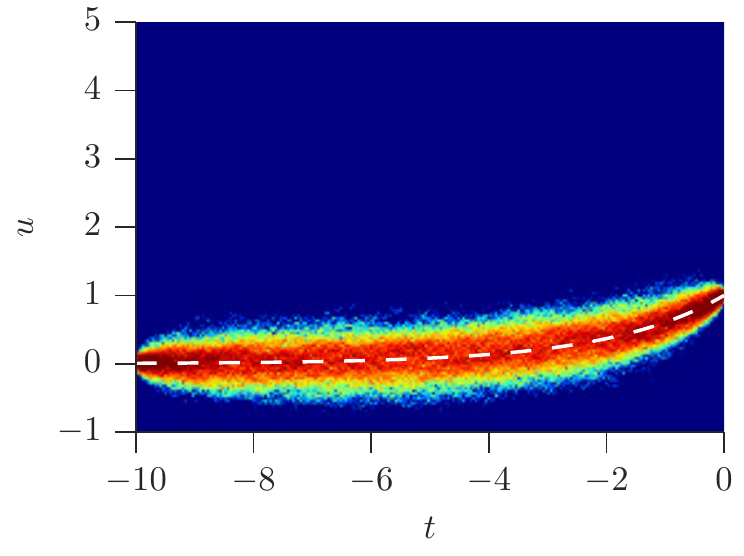} 
  \caption{Ornstein-Uhlenbeck process ($\gamma = 0.5, \sigma=0.25$):
    mean trajectory for the diffusive case compared to instanton
    prediction (left), and space-time histogram of all qualifying
    trajectories (right).}
\label{fig:OU-bm} 
\end{figure} 
How can we obtain a numerical approximation of this minimizer via
direct numerical simulations? One way is to generate an ensemble of
realizations and filter the paths that end in a small neighborhood of
the target value $u(0)=a$, for example given by the interval
$a-\epsilon < u(0) < a+\epsilon$. Since the probability distribution
will have its maximum around the minimizer, and the fluctuations
around the minimizer are (at the leading order) Gaussian, we can
expect the mean of the filtered sample paths to approximate the
minimizer. Figure \ref{fig:OU-bm} shows the result of 10$^7$
realizations. The figure on the left presents the evolution of the
mean of the filtered paths. The figure on the right shows the result
of the simulations in a slightly different way: Here, a rectangle
$[-T,0] \times [-1,5]$ was divided up in cells and the number of hits
per cell was computed. From the Freidlin-Wentzell theory, we know that
the paths will cluster around the minimizer in the weak-noise limit of
$\sigma\rightarrow 0$. An example implementation of this algorithm is
given in listings \ref{lst:wrapper} and \ref{lst:OUfilter} in the
appendix.

\subsection{Extension to jump processes}

In this section we will comment briefly on the extension of the
filtering to jump-processes, in particular related to stochastic
differential equations (SDEs) driven by symmetric alpha-stable
processes. In the context of fluid mechanics, one might think of a
system that is driven by an already non-Gaussian noise, such as the
ocean driven by wind. In general, such SDEs have attracted increasing
interest over the last decade, not only in the context of turbulence,
but also in biology and particularly in finance
\cite{cont-tankov:2004}.  An area of active research regarding fluids
is the question whether Levy walks can be used to model pair
dispersion in turbulent flows related to the Richardson law
\cite{jullien-paret-tabeling:1999,sokolov-klafter-blumen:2000}.

In the present work, we restrict ourselves to the simplest basic
approach to review how one can use path integrals in the context of
Levy-driven stochastic differential equations. Levy processes are also
called {\em fractional noise processes} as their characteristic
function (see below) involves a non-integer exponent $\alpha$. It is
well known that, using the MSRJD approach, for such Levy processes, a
generalization of the Wiener-path integral can be formulated
\cite{tarasov-zaslavsky:2008}.  The starting point of our discussion
is the stochastic differential equation (\ref{eq:sde_wentzel}) where
the Brownian increment $dW$ is now replaced by the Levy-increment
$dL$,
\begin{equation} \label{eq:sde_levy}
du = b[u]\,dt + \sigma dL\,.
\end{equation}
For simplicity, let us discuss the above equation in a one-dimensional context. For symmetric alpha-stable processes, the
characteristic function of $L_t$ is given by
\begin{equation}
\mathbb{E}({\mathrm{e}}^{isL_t}) = {\mathrm{e}}^{-t|s|^{\alpha}}\,.
\end{equation}
In the following we assume $1<\alpha<2$. The characteristic function lacks smoothness around the origin, which gives rise to fat tails. Indeed, 
it can be shown \cite{zumofen-klafter-blumen:1990} that (set $t=1$) we have
\begin{equation}
  P(u) = \frac{1}{2\pi} \int_{\mathbb{R}} {\mathrm{e}}^{-isu} \varphi(s) \,ds =  \frac{1}{2\pi} \int_{\mathbb{R}} {\mathrm{e}}^{-isu-|s|^{\alpha}}  \,ds \sim  \frac{1}{|u|^{\alpha+1}}, 
\end{equation}
as $|u|\rightarrow \infty$. While the theory of SDEs driven by Levy
processes is much more involved compared to SDEs driven by Brownian
motion, it is relatively simple to adapt the filtering technique to
such equations. Equation~(\ref{eq:sde_levy}) can be discretized in the
following form:
\begin{equation}
u_{n+1} = u_{n} + b(u_n)dt + r_n dt^{1/\alpha}\,,
\end{equation}
where the random number $r_n$ is generated, for example, by a Box-Muller-like algorithm: Draw $v_n$ from
a uniform distribution on $(-\pi/2,\pi/2)$ and draw $w_n$ from an exponential distribution with mean 1. Then compute
\begin{equation}
r_n = \frac{\sin(\alpha v_n)}{(\cos(v_n))^{1/\alpha}}\left(\frac{\cos(v_n-\alpha v_n)}{w_n}\right)^{(1-\alpha)/\alpha}\,.
\end{equation}
This simulation technique can be extended to multi-dimensional
processes in a straight forward way, using a Cholesky decomposition in
order to account for a specific correlation function. Even in simple
cases, however, the behavior of the solution differs fundamentally
from the diffusive case. In order to illustrate these differences, as
a simple example, we will discuss a Levy-driven Ornstein-Uhlenbeck
process. Note that, for such processes, the corresponding
path-integrals can be explicitly evaluated
\cite{janakiraman-sebastian:2012}. For our numerical simulations, we
have the Ornstein-Uhlenbeck process $u$ follow the SDE given by
\begin{equation} \label{eq:sde_levy_OU}
du = b(u)\,dt + \sigma dL\,, \qquad b(u) = -\gamma u\,.
\end{equation}
Here, again, we discuss the exit from a stable fixed point at negative infinity such that $u(t=0) = a$. For the Levy case, consider as an example $\alpha = 1.8$. Applying the filtering technique, we see that the filtered mean $\langle u \rangle $ of the process for the Levy case follows a similar form given by
\begin{equation} \label{eq:mean_OU}
  \langle u \rangle \approx a\,{\mathrm{e}}^{\gamma (\alpha -1) t}\,.
\end{equation}
\begin{figure} [tb]
  \centering
  \includegraphics[width=0.48\textwidth]{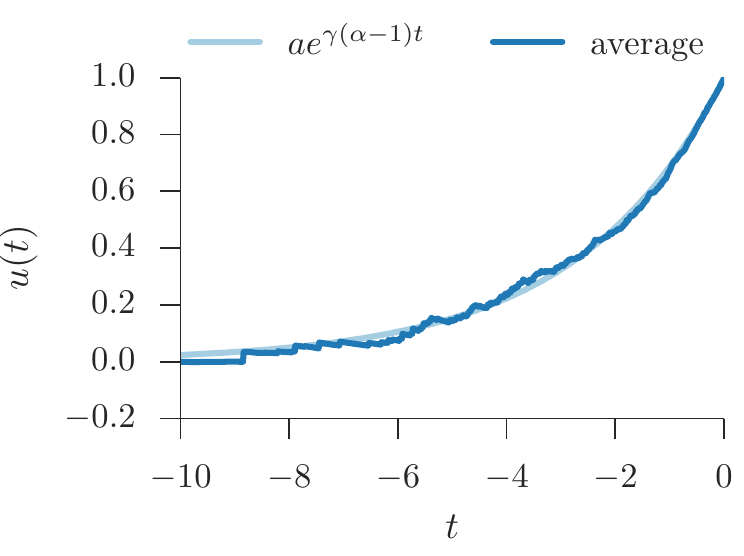} 
  \hfill
  \includegraphics[width=0.48\textwidth]{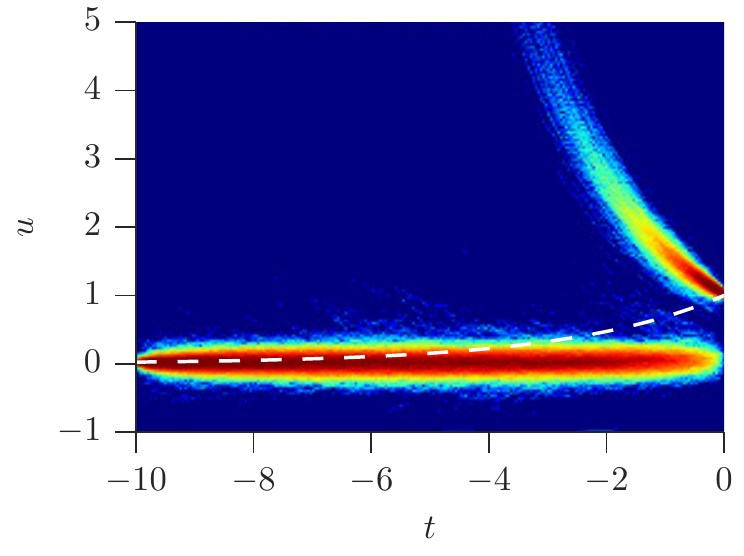} 
  \caption{Mean and trajectories for the Ornstein-Uhlenbeck ($\gamma =
    0.5, \sigma=0.08$) for the $\alpha$-stable case ($\alpha=1.75$):
    mean trajectory to instanton prediction (left), and space-time
    histogram of all qualifying trajectories (right).}
  \label{fig:OU-levy} 
\end{figure} 
Figure \ref{fig:OU-levy} shows again the result of $10^7$
simulations. As for the diffusive case, the figure on the left
presents the evolution of the mean of the filtered paths and the
figure on the right shows the histogram for a rectangle $[-T,0] \times
[-1,5]$. From here we see that, although the mean $\langle u \rangle$
appears to yield a smooth curve, the typical exit path exhibits an
entirely different evolution: It deviates only very little from the
stable fixed point at $u=0$ until a (random) time $t_r$, only to then
perform a jump of height $a\,{\mathrm{e}}^{\gamma t_r}$, such that the
deterministic drift $b$ will take it to the point $a$ at the time
$t=0$. In the diffusive case, as we have seen above, the filtered
paths were distributed around the mean paths which coincided with the
minimizer of the Freidlin-Wentzell action. An example implementation
for the case of Levy noise given in listings \ref{lst:wrapper} and
\ref{lst:Levyfilter} in the appendix.

\subsection{Filtering for the stochastic Burgers equation}

We now describe how to apply filtering in the context of the
stochastic Burgers equation. Here, the situation is more expensive
than for the Ornstein-Uhlenbeck process since we are dealing with many
more dimensions: For each realization we need to solve equation
(\ref{eq:stochastic_burgers}) with the appropriate right-hand
side. The basic idea of the filtering remains the same: We generate an
enormous number of stochastic simulations and filter the realizations
that produce the rare event that our observable measures. In the case
under consideration here, this observable is the slope of the velocity
field and we are in particular interested in the probability of large
negative gradients.

There are a variety of numerical methods available to generate the
noise term on the right hand side of equation
(\ref{eq:stochastic_burgers}) in direct numerical simulations, often
the preferred methods are Fourier-based
\cite{chekhlov-yakhot:1995-b,gotoh:1999}. The idea is that the main
results, for instance in our case the scaling of the probability
distribution of the velocity gradient, should show universal behavior
and not depend too much on the particular choice of the noise term as
long as the noise is sufficiently weak and does not dominate the
system. In the following, however, we discuss the results of a
particular numerical experiment aimed at showing the relevance of
instantons \cite{grafke-grauer-schaefer:2013}. Here, we are not only
interested in the scaling of the probability distributions but in the
detailed structure of the instantons, in particular their time
history. For this purpose, it is desirable to generate noise that
imitates the fluctuations assumed in the instanton analysis as closely
as possible. This can be done as follows:
\begin{enumerate}
\item Draw a vector $r$ of appropriately scaled normally distributed random numbers. The size of
the vector corresponds to the discretization in $x$.
\item Multiply this vector by a matrix $A$ resulting from the Cholesky
decomposition of the (discretized) correlation matrix $C$. Note that
naive discretization of $\chi$ is likely to lead to a $\tilde C$ that, due to
finite machine-precision, is not positive-semidefinite. This is due to the fact that, for a large number of discretization points, the rows of $\tilde C$ are
almost linearly dependent. In this case it is possible to
use the algorithm introduced in \cite{qi-sun:2006} in order to obtain
a matrix $C$ that is positive-semidefinite and sufficiently close to
$\tilde C$. 
\end{enumerate}
When implementing the direct numerical simulations, the available computational hardware needs
to be used in an efficient way. In our work from \cite{grafke-grauer-schaefer:2013}, 
we used a combination of accelerator graphics
cards (CUDA) and multiple processors connected via MPI: For the required resolution, the size of one 
simulation was small enough to fit on a graphics card such that a single realization 
could be performed directly on the card. In addition, other numerical procedures necessary to
extract the instanton (see following section) could be performed as well on the graphics
card. MPI was used in order to average over the different CUDA realizations. This
made the actual implementation of the algorithm fairly simple but efficient: Since the realizations are stochastically
independent, we obtain a linear scaling with the number of graphics cards.

This approach, in principle, is applicable to any generic stochastic partial differential equation with one spatial dimension. Already for two dimensions (see section \ref{sec:higher-D}), the memory requirements grow significantly, so that in these cases more sophisticated numerical algorithms have to be employed.

\subsection{Extracting the instanton}

In order to obtain the instanton from an ensemble of direct numerical simulations,
the following {\em filtering} can be applied: Prescribe a small interval
around the desired value of the observable and keep the relations 
for which the value of the observable falls into this interval. This is similar
to the procedure previously discussed and illustrated for the Ornstein-Uhlenbeck
process. In the concrete case for Burgers equation, we prescribe the
interval around a specified velocity gradient at $x=0$ at $t=0$. We assume 
that the realizations start at $t_{\mathrm{min}}$ from a zero initial condition 
and are integrated to the final time $t=0$.  For computational efficiency, 
our filtering algorithm does not only look at the
point $x=0$ but makes use of spatial translation invariance. This requires
shifting of both the velocity and the forcing field. 
In the simple case of the Ornstein-Uhlenbeck process discussed
above, such shifting was unnecessary since we were working in only one
dimension. In the case of a stochastic partial differential equation
like the stochastic Burgers equation, however, the rare event (here
the large negative gradient) can occur at any place in the
profile. The above described procedure for searching the maximum
gradient and shifting the fields allows for detecting many more rare
events.

The averaging procedure now consists of taking the average of all
those shifted fields $u_{\mathrm{shifted}}(t,x)$. Note that, from the
same simulations, we obtain also a corresponding force field
$\eta_{\mathrm{shifted}}(t,x)$ and, for both ensembles, we can compute
the ensemble average $\langle u_{\mathrm{shifted}}(t,x)\rangle$ and
$\langle \eta_{\mathrm{shifted}}(t,x)\rangle$ in space and time. Due
to the $\delta$-correlation of the driving noise (in time), we expect
convergence only after many realizations. This filtering approach can
be interpreted as the numerical counterpart to the MSRJD formulation
of the path integral taking into account the chosen observable
$O(u)=\langle \delta(u_x(0,0)=a)\rangle$ as the conditional
mean of the observable (where the conditioning is happening in our
case due to the constraint of the velocity gradient at the end point
of the evolution). For sufficiently strong gradients, corresponding to
sufficiently rare events, we therefore expect that the ensemble
averages $\langle u_{\mathrm{shifted}}(t,x)\rangle$ and $\langle
\eta_{\mathrm{shifted}}(t,x)\rangle$ will be close to the instanton
solution of (\ref{eq:instanton_u_msr}), (\ref{eq:instanton_mu}). The
corresponding optimal force $\langle
\eta_{\mathrm{shifted}}(t,x)\rangle$ is related to the instanton
solution of (\ref{eq:instanton_u_msr}) via $\langle
\eta_{\mathrm{shifted}}(t,x)\rangle = -i\chi\mu$.

\subsection{Comparison between typical shock events in DNS and instanton prediction}

The first important choice in the direct numerical simulation is
setting the initial time $T_{\min}$.  Clearly, from an analytical
point of view, one requires $T_{\min} \rightarrow -\infty$ and,
therefore $-T_{\min}$ should be as large as possible in order to let
the instanton develop. The maximum $\Delta t$ for the resolution in
time (usually the choice of $\Delta t$ is enforced by a stability
condition of the numerical scheme and related to the spatial
discretization) sets the number $N_t$ of discretization points in
time. We found that, in fact, the results are very sensitive to the
choice of $T_{\min}$ if $-T_{\min}$ is too small. Therefore, it is
recommended to perform several runs with a variety of values for the
parameter $T_{\min}$ in order to be sure that $-T_{\min}$ is
sufficiently large. In our simulations, we found that a decent value
of $T_{\min}$ is obtained by comparing $T_{\min}$ to the integral time
$T_L$ and we chose $|T_{\min}|>10T_L$.  The averages were obtained
from about $10^7$ stochastically independent realizations.

\begin{figure}[p]
\begin{center}
\includegraphics[width=0.48\textwidth]{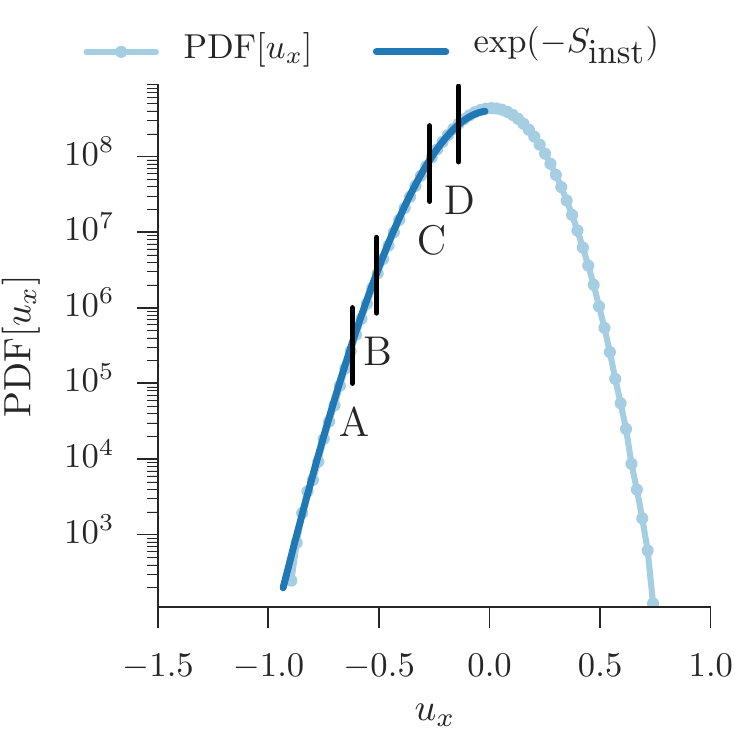} \hfill
\includegraphics[width=0.48\textwidth]{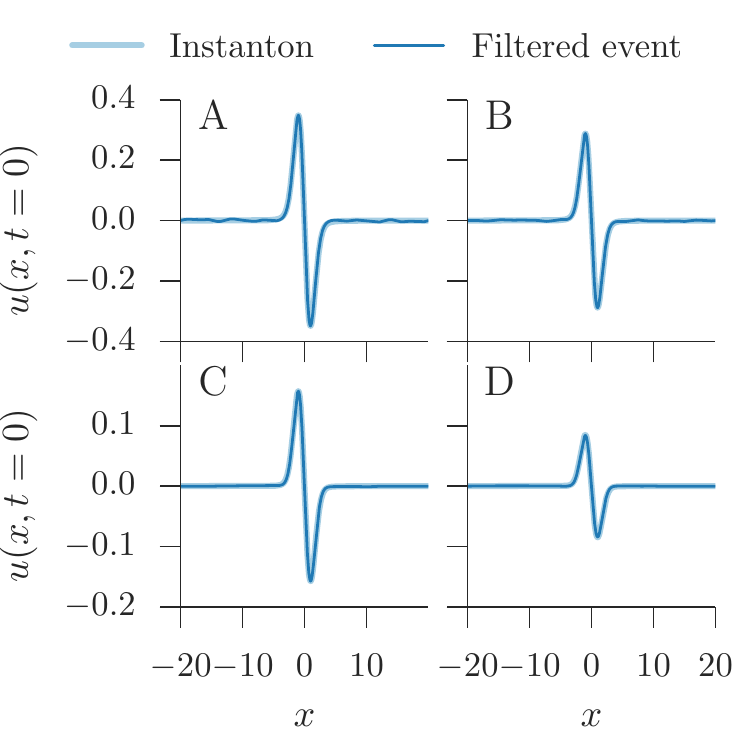}
\end{center}
\caption{Comparison between the instanton prediction and the filtered
  velocity field for the Burgers equation at a low forcing prefactor,
  where the instanton approximation becomes asymptotically
  exact. Left: The measured PDF for the velocity gradient agrees with
  the instanton prediction $\exp(-S_{\mathrm{inst}})$ over the whole
  left tail. The black vertical bars denote the gradients $u_x$ of the
  filtering comparison shown on the right. Right: The filtered events
  for gradients all over the left tail agree with the shock structure
  predicted by the instanton. \label{fig:filtering_low}}
\end{figure}

\begin{figure}[p]
\begin{center}
\includegraphics[width=0.48\textwidth]{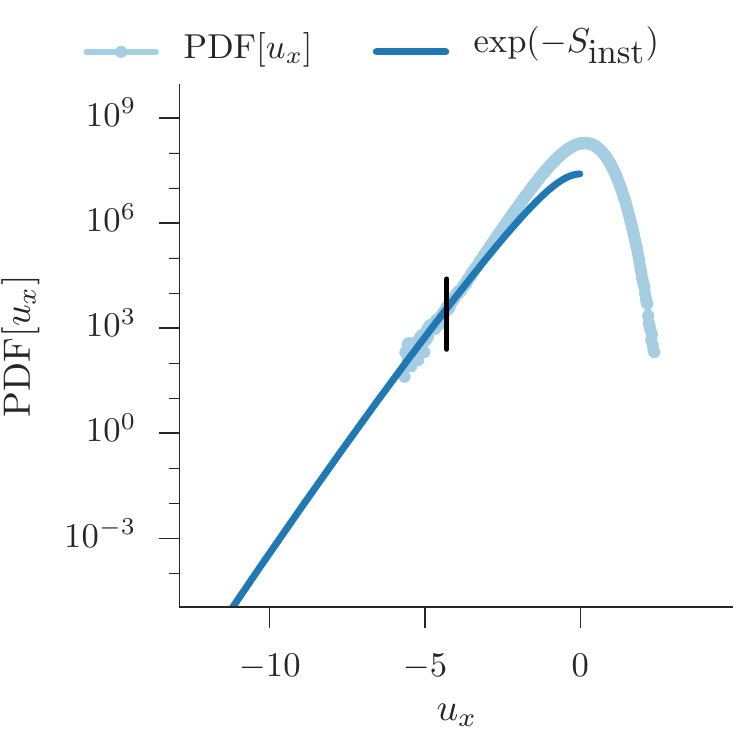} \hfill
\includegraphics[width=0.48\textwidth]{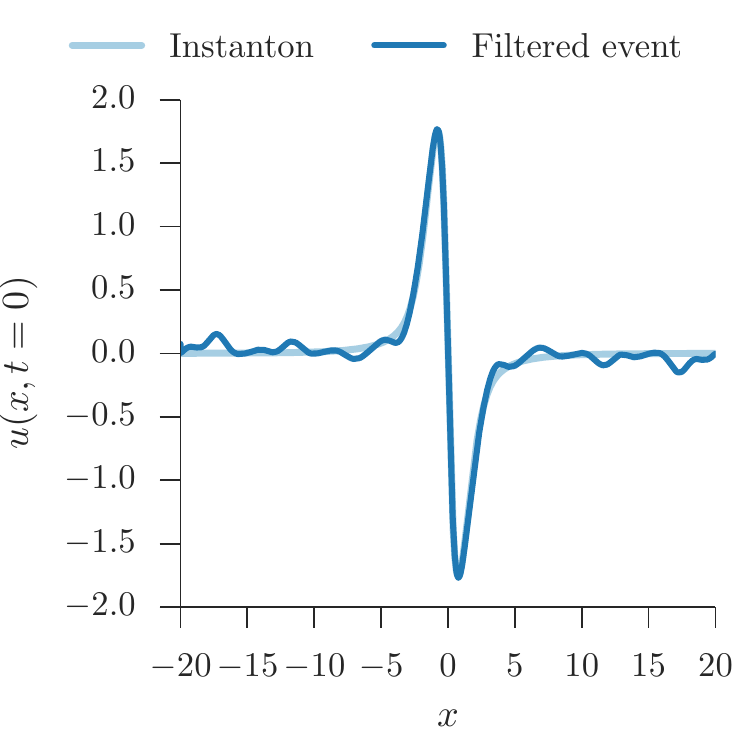}
\end{center}
\caption{Comparison between the instanton prediction and the filtered
  velocity field for the Burgers equation at a higher forcing
  prefactor, where the instanton approximation is accurate only in the
  far left tail (extreme events). Left: The scaling of the PDF tail
  agrees with the instanton prediction $\exp(-S_{\mathrm{inst}})$ for
  events of extreme negative gradient only. The black vertical bar
  denotes the gradients $u_x$ of the filtering comparison shown on the
  right. Right: The filtered event for an extreme negative gradient
  agrees with the shock structure predicted by the
  instanton. \label{fig:filtering_high}}
\end{figure}

\begin{figure}[tbh]
  \begin{center}
    \includegraphics[width=\textwidth]{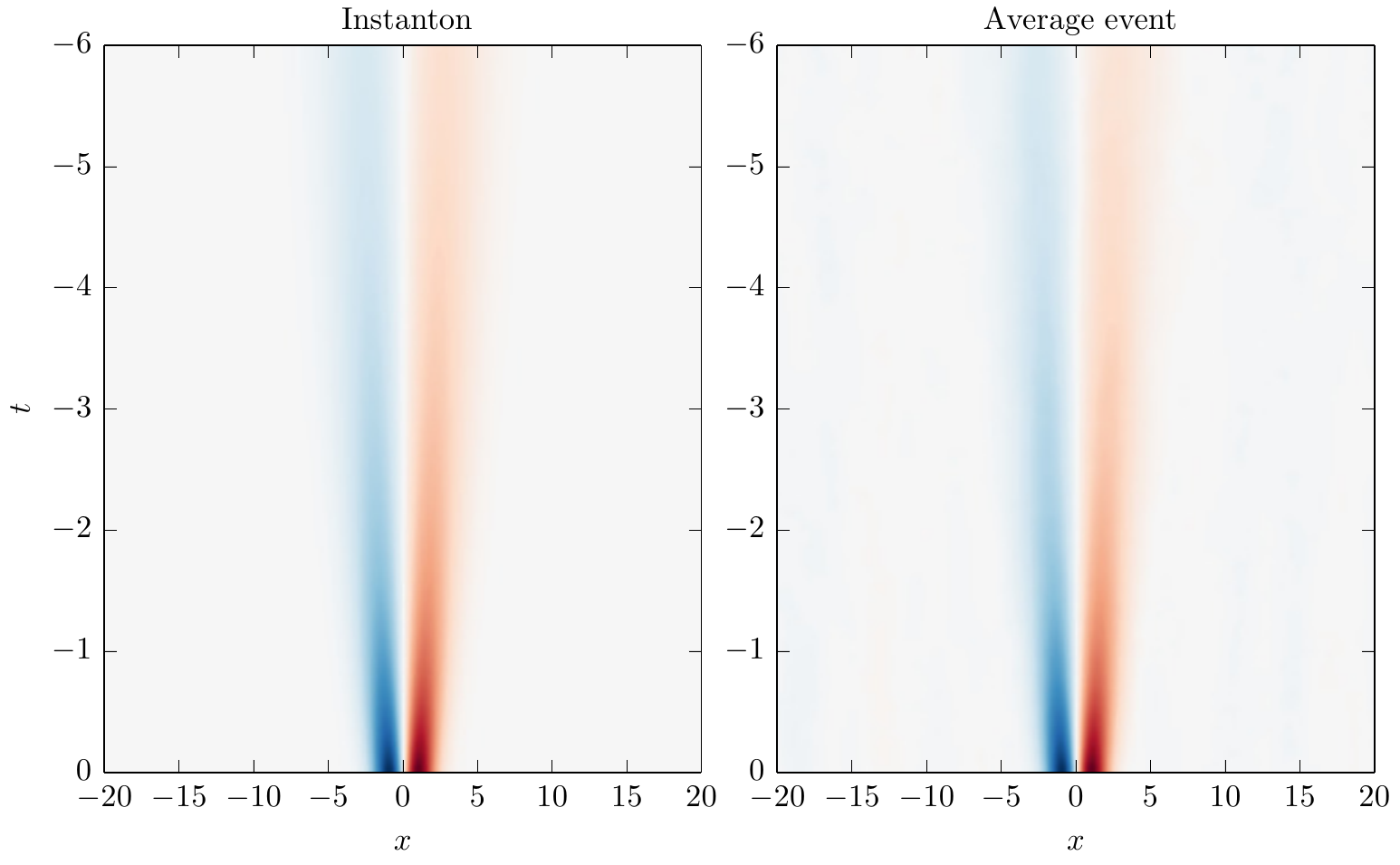}
  \end{center}
  \caption{Comparison of the velocity field between the instanton
    prediction and the average event in space-time for a moderate
    gradient, blue denoting positive and red denoting negative
    values. The time evolution of the typical shock event is
    indistinguishably reproduced by the instanton prediction. For more
    details, see
    \cite{grafke-grauer-schaefer:2013}. \label{fig:spacetime_compare}}
\end{figure}

Using this data set of simulations at different forcing strengths (or
equivalently, different Reynolds numbers), we can compare the
applicability of the instanton approximation for events of extreme
negative gradients in turbulent Burgers flows. In
figures~\ref{fig:filtering_low} (left) and~\ref{fig:filtering_high}
(left) this comparison is demonstrated for low and high values of the
forcing prefactor, respectively. The instanton prediction for the
left velocity gradient PDF tail, $\exp(-S_{\mathrm{inst}})$, agrees
over the whole range of gradients in the low forcing limit, where the
instanton approximation becomes asymptotically exact. As expected, for
a higher value of forcing, the instanton prediction captures only the
scaling of the tail for events of extreme negative
gradients. Figures~\ref{fig:filtering_low} (right)
and~\ref{fig:filtering_high} (right) on the other hand demonstrate the
agreement of the final configuration $u_\mathrm{inst}(x,t=0)$ of the
instanton against the filtered field $\langle
u_{\mathrm{shifted}}(t,x)\rangle$ for low and high values of forcing,
respectively. While in the limit of low forcing the average shock
event resembles the instanton prediction almost exactly over the whole
range of gradients, for higher values of the forcing the agreement is
only visible in the far left tail of the velocity gradient PDF.

As expected, the filtering approach does not only extract the
final configuration at time $t=0$ but also the entire time history
of the evolution of the instanton and of the optimal force which
is compared in Figure~\ref{fig:spacetime_compare} to the
solution of the instanton equations. The agreement is remarkable.

\section{Higher-dimensional problems}
\label{sec:higher-D}

Applications in fluid dynamics rarely are restricted to one spatial
dimension: The most interesting phenomena, especially in turbulence,
occur only in 2D or even 3D fluid models. Treating higher-dimensional
fluid equations numerically in the instanton framework poses a
considerable challenge though, and has rarely been attempted (notable
exceptions are the computation of transition paths for multi-stable
geophysical flows \cite{bouchet-simonnet:2009,
  bouchet-laurie-zaboronski:2011, laurie-bouchet:2015}). A major
difficulty lies in the fact that the numerical minimization of the
action functional requires the storage of the field variable and
possibly the auxiliary variable for every instance in time, which
quickly exceeds limitations of available memory. In more concrete
terms, if one is able to integrate in time a fluid equation
(e.g.~Navier-Stokes) with $N_x$ degrees of freedom (e.g. $N_x=128^3$),
one needs $N_t$ times the resources to store the corresponding
instanton trajectory, where $N_t$ is the number of time-steps of the
simulation. In the following, we will discuss modifications to the
algorithm laid out in sections \ref{sec:instanton_computation} and
\ref{sec:instanton_filtering} to efficiently compute the instanton
trajectory for higher-dimensional problems and show how to overcome
the memory restrictions.

\subsection{Recursive solution of the mixed initial/final value problem}
\label{ssec:recursive}

The instanton equations for the field variable
\eqref{eq:instanton_u_msr} and the auxiliary variable
\eqref{eq:instanton_mu} are a mixed initial/final value problem. While
the field variable is integrated forward in time, starting from an
initial condition $u_1$ at $t=-T$, the auxiliary variable is
integrated backwards in time from the final condition $p_2$ at $t=0$.
Both equations mutually depend on each other. A similar form of a
mixed initial/final value problem is encountered in a different area
of fluid dynamics, namely specific questions regarding a passively
transported scalar in a flow: analyzing where the measured density at
a given point \emph{originates} from. A method for solving the
resulting coupled equations -- one for the evolving fluid, forward in
time, and one for the passive scalar, integrated from its given final
density backwards in time -- was described in
\cite{celani-cencini-noullez:2004}. To elucidate this method, consider
a simpler system of equations on the time interval $t\in[-T,0]$ of the
form
\begin{eqnarray}
  u_t = f(u, t), \qquad u(-T) &=& u_1 \label{eq:model1}\\
  p_t = g(u, p, t), \qquad p(0) &=& p_2\label{eq:model2}\,,
\end{eqnarray}
as described in \cite{celani-cencini-noullez:2004} and which is closely 
related to the situation in control theory \cite{gunzburger:2002}. There are two
opposing ways of solving this system numerically: (i) solving
\eqref{eq:model1} forward in time starting from $u_1$ at $t=-T$ and
saving the complete evolution of $u(t)$ along the way, then
subsequently solving \eqref{eq:model2} backwards in time, using the
stored solution $u(t)$ to evaluate $g(u,p,t)$, and (ii) solve
\eqref{eq:model2} backwards in time, starting at $t=0$ from $p_2$,
while computing $u(t)$ at each timestep by integrating equation
\eqref{eq:model1} forward in time from $u_1$ at $t=-T$. Method (i) is
$\mathcal{O}(N_t)$ in both memory and computing time, which is the
worst case in memory, while method (ii) scales only $\mathcal{O}(1)$
in memory (which is optimal), while being $\mathcal{O}(N_t^2)$ in
computing time. Note that variant (ii) is only possible due to the
fact that equation~\eqref{eq:model1} is independent of the
``auxiliary'' field $p$. In the case of mutual dependence, such as the
system of instanton equations, this variant does not apply.

\begin{figure}[tb]
  \begin{center}
    \includegraphics[width=0.6\linewidth]{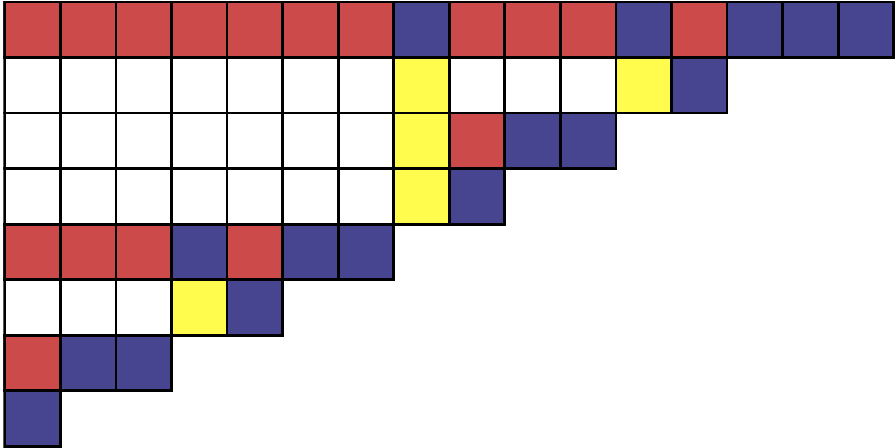}
  \end{center}
  \caption{Depiction of the recursive integration algorithm for
    $k=2$. Each square represents one of the 16 time-steps, showing
    steps computed but subsequently dropped (red), computed and
    retained in memory (blue/dark) and already stored in memory
    previously (yellow/light). Boxes are left white, if they are
    neither computed nor necessary at this point of the algorithm. The
    total memory requirement is the maximum number of green and blue
    boxes in a line (5 in this example), which is $\mathcal{O}(\log
    N_t)$. The total computation time is the total number of red and
    green boxes (33 in this example), which is $\mathcal{O}(N_t\,\log
    N_t)$. For more details, see \cite{grafke-grauer-schindel:2014}.}
  \label{fig:recursive}
\end{figure}

These two building blocks can now be employed to obtain an efficient
compromise -- also known as checkpointing in control theory 
\cite{walther-griewank:1999,wang-moin-iaccarino:2009} -- 
between the two scalings and balancing memory efficiency
and computational cost. First, split the interval $[-T,0]$ into $k$
sub-intervals. After integrating \eqref{eq:model1} once and storing
the result at the beginning of each sub-interval, we reduce the
original problem to a similar problem on a shorter domain of size
$N_t/k$. Recursively, for each of these sub-intervals, we can break
down the problem by splitting the domain, until we reach an interval
length of only a single integration step. At this point we can carry
out the integration. Inspired by similar principles in multi-grid
algorithms or the fast Fourier-transform, a natural choice is $k=2$.
In this case, the memory requirement scales as $\mathcal{O}(\log N_t)$
and computing time as $\mathcal{O}(N_t\, \log N_t)$. A schematic
depiction of the algorithm for $k=2$ and $N_t=16$ is shown in
figure~\ref{fig:recursive}. For the initial solution of the field
equation, the field is stored at the intermediate timesteps
$i\in\{8,12,14,15,16\}$, of which the least three can be used
immediately for the backwards propagating auxiliary equation. Whenever
a timestep is encountered for which the field configuration is not
stored, such as $i=13$, it is propagated forward from the last known
position (in this case $i=12$), while storing intermediate values
recursively in the same fashion.

The algorithm described above cannot be used to solve the instanton
equations~\eqref{eq:instanton_u_msr}, \eqref{eq:instanton_mu} without
modification, because in contrast to the model
problem~\eqref{eq:model1}, \eqref{eq:model2}, both equations are
mutually dependent and neither the $u$- nor the $p$-equation can be
solved without referring to the other field. As a consequence, at
least one of the fields has to be stored for each time-step. Note
though that in equation~\eqref{eq:instanton_u_msr} the auxiliary field
only acts on $u$ through a convolution with the forcing
correlation. Since in fluid dynamical applications the forcing is
usually restricted to only a few active modes, $\chi p$ is very
compact to store. In other words, the \emph{optimal forcing} $\chi p$
is the only field that needs to be kept for every timestep to be able
to integrate the complete velocity field and auxiliary field of the
instanton configuration. We remark that this approach cannot recover
the $\mathcal{O}(\log N_t)$ scaling of the original algorithm. On the
other hand, since the number of active modes of the forcing are
constant, the memory cost of the auxiliary field becomes independent
of the number of degrees of freedom in space $N_x$. In contrast to
this, the expensive storage of the field variable scales
logarithmically in time. This interplay of savings allows for the
drastic savings in memory necessary to compute higher-dimensional
instantons.

\subsection{Application: Two-dimensional Burgers equation}

\begin{figure}[tb]
  \begin{center}
    \includegraphics[width=0.48\linewidth]{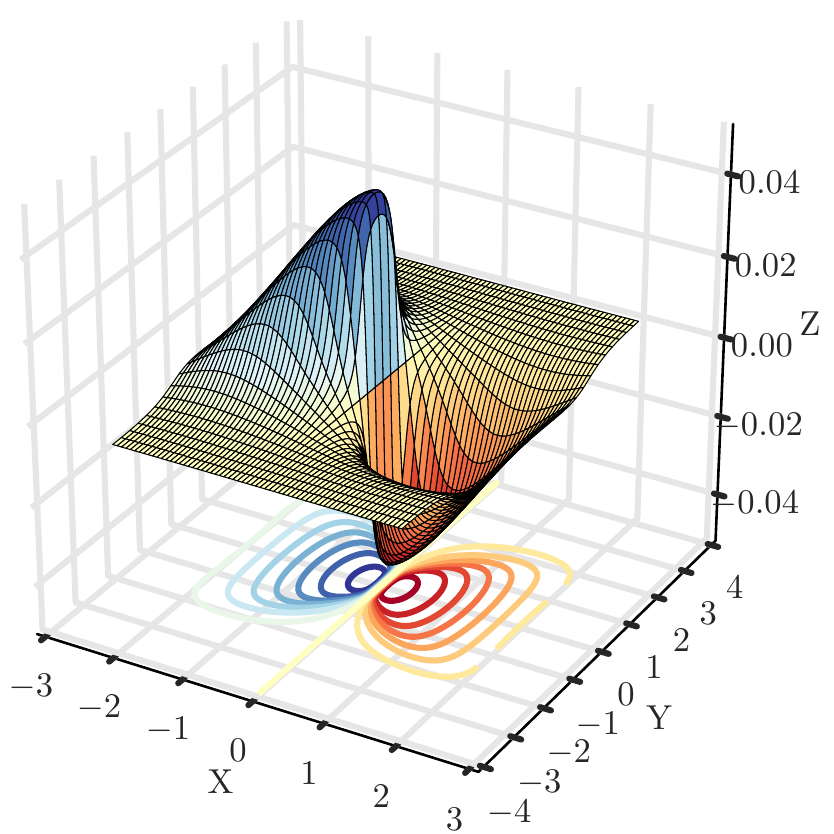}
    \includegraphics[width=0.48\linewidth]{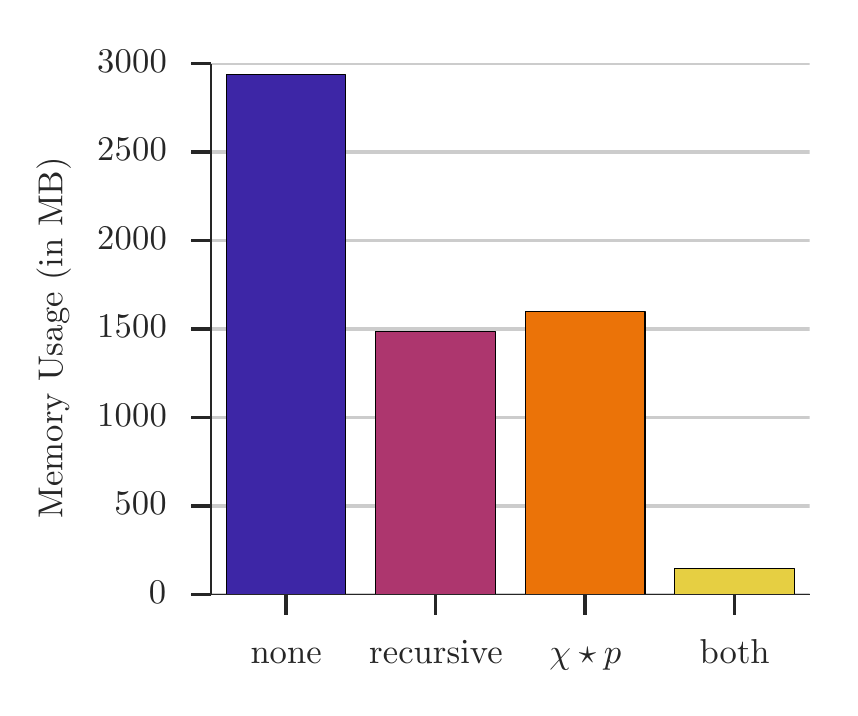}
  \end{center}
  \caption{Left: Surface plot of the $x$-component of the velocity
    field, $u_x(x,y)$, for the 2D shock instanton configuration for
    the gradient $\partial_x u_x=-1$, $\nu=10^{-2}$. Comparison of
    memory costs for 2D simulations with $N_x=256\times256$,
    $N_t=2048$ for the presented optimizations. The total memory
    saving of the combined algorithm exceeds a factor of $20$. For
    more details, see \cite{grafke-grauer-schindel:2014}.}
  \label{fig:2D-instanton}
\end{figure}

As example application for the numerical computation of
higher-dimensional instantons in fluid applications, we will consider
the Burgers equation in 2D \cite{grafke-grauer-schindel:2014},
\begin{equation}
  \label{eq:burgers-2d}
  \partial_t u + u\cdot\nabla u - \nu \Delta u = f\,,
\end{equation}
for a forcing $f(x,t)$ correlated white in time and with a finite
correlation length $L$ in space, active only on large-scale modes
(i.e. truncating the correlation function in Fourier space above a
mode $\omega_c$, as in \eqref{eq:corr_function_mod}). In contrast to
the 2D incompressible Navier-Stokes equation, for which was shown in
\cite{falkovich-lebedev:2011} that the naive instanton will be
trivial, the 2D Burgers equation exhibits a direct
cascade. Furthermore, equation \eqref{eq:burgers-2d} preserves
irrotationality of the flow under irrotational forcing, which is the
scenario we will restrict ourselves to in this section.

As generalization of the observable taken in 1D, i.e. $F[u]=\partial_x
u(x=0,t=0)$, several choices are possible: Taking $F[u] =
\nabla\cdot u$ conditions on events exhibiting extreme
velocity divergence at a single point, which corresponds to
``explosion'' and ``implosion'' events. Instead of this, we will focus
on the case
\begin{equation}
  F[u] = \partial_x u_x(0,0)\,,
\end{equation}
which selects events of high velocity gradient in a single
direction. This observable is closely related to the energy
dissipation, $\nu |\nabla u|^2$, but additionally breaks
rotational symmetry. We will identify the instanton solution with
shock structures known from 2D Burgers turbulence. The instanton
equations corresponding to equation \eqref{eq:burgers-2d} read
\begin{eqnarray}
  \partial_t u + u \cdot \nabla u - \nu \Delta u &=& \chi p\\
  \partial_t p + u \cdot \nabla p - (p \times \nabla) u^\perp + \nu \Delta p &=& 0\,,
\end{eqnarray}
with $u^\perp = (-u_y, u_x)$ and the convolution $(\chi p)_i = \sum_j
\chi_{ij} p_j$. Figure \ref{fig:2D-instanton} (left) depicts the
solution of these equations with the numerical scheme laid out
above. Shown is a surface-plot of the $x$-component of the velocity
field for the instanton around the origin at $t=0$. As expected, the
size of the shock structure across the shock is determined by the
viscosity $\nu\ll1$, while its length along the shock is given by the
forcing correlation length $L=1$.

This example demonstrates the drastic savings in memory that are
possible by employing the recursive technique: The highest resolution
achieved for the 2D Burgers problem is $N_x=1024\times1024$,
$N_t=8192$, with a total memory usage of $577$MB. This fits completely
into the memory of a single graphics card on which the computation was
undertaken. In contrast, the extrapolated memory usage of the
un-optimized algorithm would be about $300$ times higher. Note also
that the computational overhead of the optimization, due to
logarithmic scaling, is slightly lower than a factor $3$ in the total
computation time. Figure~\ref{fig:2D-instanton} (right) shows a
comparison of the memory cost for a lower resolution setup of
$N_x=256\times256$, $N_t=2048$ in order to fit the unoptimized case on
the machine: Both the recursive optimization and the projection method
amount to a saving of roughly one half, the combination of both leads
to memory savings of a factor $20$.

\subsection{Application: Three-dimensional Navier-Stokes equations}

\begin{figure}[tb]
  \begin{center}
    \includegraphics[width=0.7\linewidth]{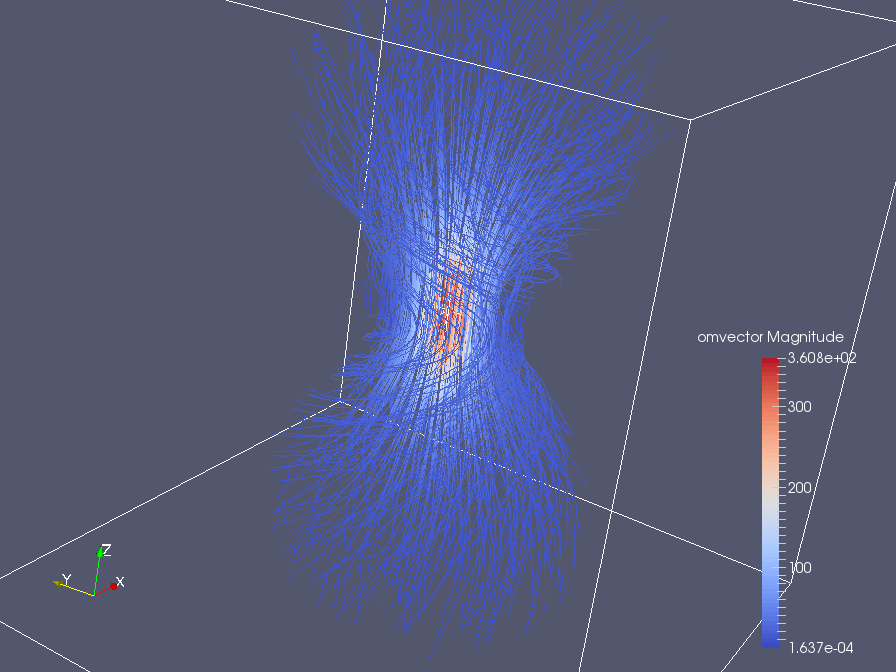}
  \end{center}
  \caption{Filtered vorticity field of a turbulent 3D incompressible
    Navier-Stokes flow.}
  \label{fig:3D-instanton-lines}
\end{figure}

\begin{figure}[ptb]
  \begin{center}
    \includegraphics[width=0.98\linewidth]{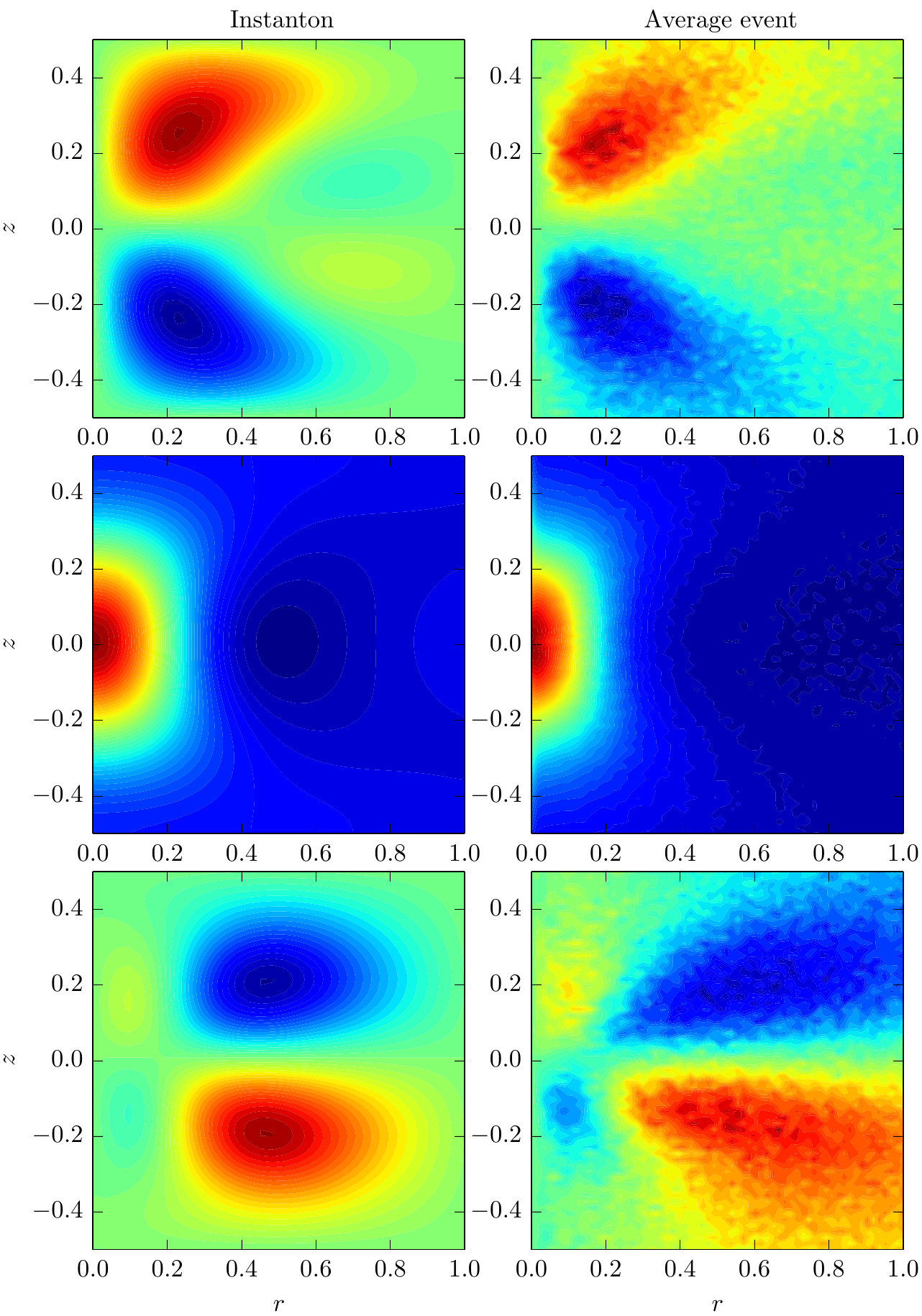}
  \end{center}
  \caption{Comparison of filtered vorticity field (right) versus
    instanton solution (left) for the 3D incompressible Navier-Stokes
    equation in cylindrical coordinates. All three vorticity
    components are shown, from top to bottom: $\omega_r$, $\omega_z$,
    $\omega_\theta$.}
  \label{fig:3D-instanton}
\end{figure}

As last example of higher-dimensional fluid systems to be treated with
the instanton formalism we will consider the three-dimensional
incompressible Navier-Stokes equations,
\begin{equation}
  \label{eq:navier-stokes}
  \partial_t u + u \cdot \nabla u - \nabla p -\nu \Delta u = f, \quad \nabla\cdot u=0\,.
\end{equation}
This equation of course lies at the very center of turbulence
research. In spirit of the program laid out above there is hope to
link the dissipative structures of turbulent flows to the
configuration obtained by the instanton formalism. This in turn might
elucidate not only the form and emergence of such structures, but
serve to answer questions about the intermittent nature of turbulent
flows. Yet, it is far from obvious that the naive approach of
formulating the corresponding MSRJD-action \eqref{eq:action_MSR} for
the Navier-Stokes equations \eqref{eq:navier-stokes} and then
computing its minimizing trajectory (which in itself is a considerable
amount of work) leads to any meaningful results: It is not clear which
observable to consider, a choice which can change the structure of the
instanton trajectory completely. One might be tempted to look at
events of high gradients (like in Burgers) or high energy dissipation
\begin{equation}
  F[u] = \nu |\nabla u(x=0,t=0)|^2
\end{equation}
to obtain the dissipative structures of turbulence. Yet, these events
might be well into the dissipative range and thus not capture the
multi-fractal nature of the turbulent cascade in the inertial range.

On the other hand, one can obviously commence on the path laid out in
sections \ref{sec:instanton_computation} and
\ref{sec:instanton_filtering} for the Burgers equation: Comparing the
events dominating the extreme tails of the PDF of a chosen observable
to the structures obtained by solving the corresponding instanton
equations. This way, one gets access to the tail scaling of PDFs of
turbulent quantities that are much harder to obtain by classical
methods. In order to demonstrate that such an approach is feasible, we
show preliminary results of a numerical computation of a 3D
Navier-Stokes instanton in figure \ref{fig:3D-instanton-lines}. As
observable,
\begin{equation}
  F[u] = \nabla\times u(x=0,t=0)
\end{equation}
was chosen, which amounts to events of extreme vorticity in the origin
at $t=0$. The resulting structure in quality resembles the well-known
vorticity filaments observed in developed 3D Navier-Stokes
turbulence. For the filtered fields, the sample size is $\approx
10^3$.

Structures at the final time $t=0$ are very similar to those first obtained by Novikov
\cite{novikov:1993, mui-dommermuth-novikov:1996} and analyzed in
detail by Wilczek \cite{wilczek:2011}. They
consider the conditionally averaged vorticity field
\begin{equation}
  \omega_\mathrm{filter} = \langle \omega(x,t) | \omega(x=0,t=0)=a \rangle\,,
\end{equation}
i.e. the vorticity field in space and time, conditioned on the fact
that it will attain the vorticity $\omega(0,0)=a$ in the
course of its evolution. This is of course the same as the
``filtering''-approach laid out above for Burgers shock structures. It
therefore makes sense to ask whether the instanton resembles the
filtered structure for events of extreme vorticity. Preliminary
results of this are shown in figure \ref{fig:3D-instanton}. Due to the
symmetry of the observable, the instanton solution observes rotational
symmetry around the axis prescribed by the vorticity in the
origin. The filtered velocity field is obtained by averaging fields of
the same vorticity $\omega(x,t)=a$, after translating and rotating
them into the origin. Depicted is a direct comparison of all three
components of the vorticity field in cylindrical coordinates
$(r,\theta,z)$ between the instanton configuration and the filtered
vorticity field of a turbulent DNS of the 3D Navier-Stokes equation,
both with $N_x=128^3$ and $N_t=128$ for the instanton. The instanton
field reproduces the conditionally averaged one remarkably well. In
particular, features like the spreading of the vortex and the
appearance of a swirl for larger radii (also visible in figure
\ref{fig:3D-instanton-lines}) is well-reproduced. These results serve
as evidence that the techniques laid out above are powerful enough to
compute instantons even for the 3D setup.

In how far this Navier-Stokes vorticity instanton can be analytically
captured by the approximation derived in \cite{moricone:2004} remains
a future task.

\section{Outlook}

Instantons (saddle point configurations) of functional integrals,
equivalent to minimizers of the associated Freidlin-Wentzell action,
govern the tails of probability distributions of physically relevant
observables in stochastically driven systems.  In this work, we
presented several algorithms to compute such instantons efficiently --
either by extracting them from direct numerical simulations via
filtering or by iteratively solving the associated Euler-Lagrange
equations.  For the stochastically forced Burgers equation, agreement
of both methods was shown in one and two dimensions. Preliminary
results for the three-dimensional Navier-Stokes equations are
encouraging, but much higher resolutions are needed to see whether
both approaches will also lead to similar results in this case. This,
however, is only a first step on a long path: As it is well-known from
quantum field theory, fluctuations around the instanton can yield to
non-trivial modifications of the results and are, quite often, of
major relevance for understanding the underlying physics of the
system. We expect this to hold for fluids as well, meaning that an
appropriate and efficient computational framework to capture the
impact of fluctuations needs to be developed. Moreover, at this stage,
the physical role of the found instantons needs to be investigated
further: What is their impact (if any) on intermittency and the
scaling of structure functions in the inertial range? Can they indeed
capture the impact of singular structures in real flows in a variety
of settings (not only in Burgers, but also in Navier-Stokes, MHD,
(surface) quasi-geostrophic, etc.)? What if the underlying PDE is
integrable (or has at least solitary wave solutions) -- can these
properties be exploited from a computational point of view? We believe
that instantons are relevant for a wide range of phenomena in fluid
dynamics -- and finding them computationally and understanding their
nature is still an area of on-going and fruitful research.

\ack

We would like to thank Stephan Schindel for his numerical work, in
particular regarding the 3D instanton computations, and Holger Homann
for providing figure \ref{fig:dissipative-structures}. We thank
Gregory Falkovich, Eric Vanden-Eijnden, Maxim Polyakov and Freddy
Bouchet for helpful discussions.
We also would like to thank an anonymous referee and Pierre Hohenberg for
pointing out important literature concerning the historical
development of the instanton/optimal fluctuation approach.
The work of T.G. was partially supported through the grants
ISF-7101800401 and Minerva-Coop 7114170101. The work of R.G. 
benefited from partial support through DFG-FOR1048, project B2. The
work of T.S. was partially supported by the NSF grant DMS-1108780.

\section*{References}

\bibliographystyle{unsrt}
\bibliography{bib}

\definecolor{mygreen}{rgb}{0,0.6,0}
\definecolor{mygray}{rgb}{0.5,0.5,0.5}
\definecolor{mymauve}{rgb}{0.58,0,0.82}
\lstset{language=Python, numbers=left, %
        basicstyle=\footnotesize\ttfamily, %
        commentstyle=\itshape\color{mygreen},    
        keywordstyle=\color{blue},       
        numberstyle=\tiny\color{mygray}, 
        stringstyle=\color{mymauve},     
        rulecolor=\color{black},         %
        tabsize=2,                       
        showstringspaces=false,          
        breaklines=true,                 
}

\appendix
\section{Example source code}

\begin{lstlisting}[caption={Solution of the instanton equation using the
    Chernykh-Stepanov scheme for the Ornstein-Uhlenbeck process, with
    $b(u)=-u-\frac12u^3$. Compare to the analytical solution $u(t) =
    \sqrt{2/(3\exp(-2t)-1)}$.},label={lst:OUiter}]
import numpy as np
import pylab

def iterativeInstantonEquations():
    pEnd = 3; N=1000;
    eta = 1; iterations=200;
    t = np.linspace(-10,0,N); dt = t[1]-t[0];

    x = np.zeros((N,1))
    p = np.zeros((N,1))
    p[-1] = pEnd

    for it in range(iterations):
        p_ = integrate_P(x,p,N,dt)
        x_ = integrate_X(x,p,N,dt)
        x = eta*x_ + (1.-eta)*x
        p = eta*p_ + (1.-eta)*p
        
    pylab.plot(t, x, 'x-b')
    pylab.plot(t, p, 'x-r')
    pylab.show()

    return [x, p]

def integrate_X(x, p, N, dt):
    ret = x; xR = x[0];
    for i in range(N-1):
        xR = xR + dt*(p[i]+b(xR))
        ret[i+1] = xR
    return ret

def integrate_P(x, p, N, dt):
    ret = p; pR = p[N-1];
    for i in range(1,N):
        pR = pR + dt*bPrime(x[N-i])*pR
        ret[N-i-1] = pR
    return ret

def b(x):
    return -x-0.5*x**3

def bPrime(x):
    return -1-1.5*x**2
\end{lstlisting}

\begin{lstlisting}[caption={Wrapper script with standard parameters and
    plotting functions. Will produce plots like those in figures
    \ref{fig:OU-bm} and \ref{fig:OU-levy}. Uncomment line 26 to switch
    between Gaussian noise and Levy
    noise.},label={lst:wrapper}]
import numpy as np
import pylab

def plotAverage(t, paths, params):
    # Plot average filtered path, and prediction
    pylab.plot(t, np.mean(paths,0), 'b-')
    pylab.plot(t, params["a"]*np.exp(params["k"]*t), 'r--')
    pylab.show()

def plotHistogram(t, paths, params):
    # Plot space-time histogram
    instHist = np.zeros((params["nbins"], params["Nt"]))
    bins = np.linspace(-2,2,params["nbins"]+1)
    for i in range(params["Nt"]):
        h,b = np.histogram(paths[:,i], bins=bins)
        instHist[:,i] = h

    pylab.imshow(np.flipud(instHist), extent=(params["tDomain"][0], params["tDomain"][1], 2, -2))
    pylab.show()

def start():
    params = {"nPaths":1e4, "Nt": 100, "nbins": 100, "tDomain": (-10,0), 
              "a": 1.0, "k": 1.0, "epsilon": 0.1, "sigma": 0.4}

    t, paths = filterGauss(params)  # Gaussian force
    #t, paths = filterLevy(params)  # Levy force

    pathCount = paths.shape[0]
    print("Found {} trajectories ({}%)".
          format(pathCount, np.round(pathCount/params["nPaths"]*100.0,4)))

    plotAverage(t, paths, params)
    plotHistogram(t, paths, params)
\end{lstlisting}

\begin{lstlisting}[caption={Filtering the instanton trajectory for an
    Ornstein-Uhlenbeck process.},label={lst:OUfilter}]
def filterGauss(params):
    sigma = params["sigma"]
    k = params["k"]
    t = np.linspace(params["tDomain"][0], params["tDomain"][1], params["Nt"])
    dt = t[1]-t[0]
    sigsqrtdt = params["sigma"]*np.sqrt(dt)
    u = np.zeros((params["nPaths"], params["Nt"]))

    # Integrate Paths in parallel
    for i in range(1,params["Nt"]):
        r = np.random.randn(params["nPaths"])
        u[:,i] = (1-k*dt)*u[:,i-1]+r*sigsqrtdt

    # Return paths filtered to hit around a
    return t, u[np.abs(u[:,-1]-params["a"]) < params["epsilon"],:]
\end{lstlisting}

\begin{lstlisting}[caption={Filtering the instanton trajectory for an
    Ornstein-Uhlenbeck process, but with Levy noise.},label={lst:Levyfilter}]
def filterLevy(params):
    mu = params["mu"]
    sigma = params["sigma"]
    k = params["k"]
    t = np.linspace(params["tDomain"][0], params["tDomain"][1], params["Nt"])
    dt = t[1]-t[0]
    sigdtm = params["sigma"]*dt**(1/mu)
    u = np.zeros((params["nPaths"], params["Nt"]))

    # Integrate Paths in parallel
    for i in range(1,params["Nt"]):
        v = -np.pi/2 + np.pi*np.random.rand(params["nPaths"])
        w = -np.log(np.random.rand(params["nPaths"]))
        xi = np.sin(mu*v)/(np.cos(v)**(1/mu))*(np.cos(v-mu*v)/w)**((1-mu)/mu)
        u[:,i] = (1-k*dt)*u[:,i-1]+xi*sigdtm
        
    # Return paths filtered to hit around a
    return t, u[np.abs(u[:,-1]-params["a"]) < params["epsilon"],:]
\end{lstlisting}

\end{document}